\documentclass[useAMS,usenatbib]{mn2e}

\usepackage{graphicx}
\usepackage{mathtools}
\usepackage[usenames,dvipsnames]{color}

\newcommand\be{\begin{equation}}
\newcommand\bea{\begin{eqnarray}}
\newcommand\ee{\end{equation}}
\newcommand\eea{\end{eqnarray}}
\newcommand\bw{\begin{onecolumn}}
\newcommand\ew{\end{onecolumn}}

\newcommand{\nn}{\nonumber}

\newcommand{\p}{\partial}

\title[Geodesics and moments in ST gravity]{Geodesic properties in terms of multipole moments in scalar-tensor theories of gravity}
\author[G.~Pappas and T.P.~Sotiriou]{George Pappas$^{1}$\thanks{E-mail:
Georgios.Pappas@nottingham.ac.uk} and Thomas P.
Sotiriou$^{1,2,3}$\thanks{E-mail: Thomas.Sotiriou@nottingham.ac.uk}\\
$^{1}$School of Mathematical Sciences, The University of Nottingham,
University Park, Nottingham NG7 2RD, UK\\
$^{2}$School of Physics and Astronomy, The University of Nottingham,
University Park, Nottingham NG7 2RD, UK\\
$^{3}$Perimeter Institute for Theoretical Physics, Waterloo, Ontatio, N2L 2Y5, Canada }

\begin{document}

\date{}

\pagerange{\pageref{firstpage}--\pageref{lastpage}} \pubyear{2002}

\maketitle

\label{firstpage}

\begin{abstract}
The formalism for describing a metric and the corresponding scalar in terms of multipole moments has recently been developed for scalar-tensor theories. We take advantage of this formalism in order to obtain expressions for the observables that characterise geodesics in terms of the moments. These expressions provide some insight into how the structure of a scalarized compact object affects observables. They can also be used to understand how deviations from general relativity are imprinted on the observables. 
\end{abstract}


\section{Introduction}

Low mass X-ray binaries (LMXBs) are astrophysical systems that host black holes (BHs) or neutron stars (NSs) with a regular star as a companion. The presence of a compact objects in the binary makes it a natural laboratory for some of the most extreme aspects of physics. In particular, the systems that harbor NSs involve both strong gravity effects and effects associated to the properties of matter in supra-nuclear densities. Therefore, studying these systems can broaden both our understanding of gravity, by testing the predictions of the established theory of General Relativity (GR), as well as our understanding of the equation of state (EoS) of matter for densities as high as the densities that can be found at the centre of NSs. 

There are several astrophysical observables that can be associated to the properties of the spacetime and more specifically to the properties of geodesics around the compact object that is part of such a system. 
For example, quasi-periodic oscillations (QPOs) of the X-ray flux observed from the accretion disc in LMXBs (for reviews see \cite{lamb,derKlis}) is one such phenomenon. The proposed mechanism for explaining it is the relativistic precession of geodesics (see \cite{stella}). In this context, one assumes that the observed QPOs come from the frequencies associated to circular geodesic orbits and their perturbations. The main frequencies related to these orbits are: the orbital frequency of the circular motion $\Omega$; the radial oscillation of a slightly elliptic orbit that has frequency $\kappa_{\rho}$ and the corresponding precession frequency $\Omega_{\rho}\equiv\Omega-\kappa_{\rho}$ (periastron precession); and finally the vertical oscillation of a slightly off-equatorial orbit which has frequency $\kappa_z$ and the corresponding precession frequency $\Omega_z\equiv\Omega-\kappa_z$ (nodal precession). As is customary in the relevant literature, the term ``frequency" is used loosely to refer to what is actually an angular velocity  that is equal to $2 \pi \nu$, where $\nu$ is the frequency expressed as cycles in the unit of time. 

The radiated spectrum of an accretion disc can be also associated to the properties of the geodesics around a compact object. It is associated to the energy per unit mass $E/\mu$ of a particle that has a circular equatorial orbit and how that is distributed from the outer parts of the disc down to the location of the innermost stable circular orbit (ISCO), which is essentially the position where the accretion disc terminates. Under the assumption of a thin, radiatively-efficient accretion disc, a fluid element of unit mass in the disc radiates as it moves inwards in a way that depends on the distribution of energy on the circular geodesics. This information is imprinted on the temperature distribution of the disc and its spectrum. This type of properties of an accretion disc and their connection to the background geometry have been used in the past to determine the rotation of black holes (see for example \cite{narayan} and references therein), where a technique called continuum-fitting method is applied in order to determine the spin of the black hole from the luminosity of the disc. 
   
One would like to extract information about the structure of a compact object from observables by studying the properties of the geodesics around it. One way to do so is to explore the relation between these observables and the multipole moments of the spacetime as they were defined by \cite{Geroch70I,Geroch70II,hansen,fodor:2252}. This has been the topic of extensive study in GR. \cite{Ryan95} studied the relation between the orbital and precession frequencies and the multipole moments of the spacetime. He also studied the relation between the multipole moments and the energy per unit mass, $E/\mu$, distribution of circular geodesics. In particular, he expressed the energy change per logarithmic orbital frequency change, 
\be
\frac{\Delta E}{\mu}\equiv-\frac{1}{\mu}\frac{dE}{d\ln\Omega}\,,
\ee
as well as the ratios of the precession frequencies over the orbital frequency, $\Omega_{\rho}/\Omega$ and $\Omega_z/\Omega$, as functions of the orbital frequency $\Omega$ in the form of series expansions where the coefficients of the expansions depend on the multipole moments. 

These relations were intended to be used in gravitational waves detection in order to extract the multipole moments of the spacetime around supermassive BHs, but they have proven to be useful in other ways as well. They were used to calculate the moments of numerical spacetimes by \cite{poisson} and \cite{pappas-apostolatos} and it seems that the method has become a common practice after that (with the most recent examples the application to anisotropic stars by \cite{YYanisotropic} and to tidally deformed objects by \cite{PaniTidal}). They were also proposed as a possible way of measuring the first three multipole moments of a NS from QPOs by \cite{PappasQPOs}. They have even been used to study floating orbits around BHs by \cite{KapadiaFloating}.
   
\cite{SSisco} have studied how the ISCO radius, $R_{\textrm{ISCO}}$, and the orbital frequency, $\Omega_{\textrm{ISCO}}$, at the location of the ISCO, depend on the multipole moments of the spacetime, with the aim to impose constrains on the properties of the central compact object from measurements of these quantities.  

There is no doubt, therefore, that this sort of relations between orbital properties and multipole moments have been a very useful tool in studying compact objects in GR. As a next step, one could attempt to derive similar relations for alternative theories of gravity. The motivation for this is actually 2-fold. One can obviously use such relation in the same applications that they have already been used in GR. In addition to this though, one could also exploit them in order to understand the degeneracies in the effect that deviations from GR and uncertainties in the description of the matter content of the star itself can have on actual observables.  

Recently, the multipole moments for scalar-tensor theories of gravity with a massless scalar field have been defined by \cite{PapSot:STmoments}. This opens the way for relating geodesic properties that are associated to astrophysical observables, such as frequencies, accretion spectra, and ISCO radii, to the multipole moments in  scalar-tensor theory. These relations  and the way that they compare to the corresponding GR relations, can be a very useful tool for testing GR against observations and constraining the parameters of scalar-tensor theory.

In what follows, we first give in Sec.~\ref{sec:STintro} a brief description of scalar-tensor theory with a massless scalar field, a description of stationary axisymmetric spacetimes in this theory, and the properties of circular geodesics in these spacetimes. In Sec.~\ref{sec:spherical} we calculate expressions for $\Delta E/\mu$, the various frequencies, and the ISCO radius in terms of the moments for the spherically symmetric solution in scalar-tensor theory. This serves as a first demonstration of the methods that we then use in Sec.~\ref{sec:generalST} in order to calculate the same expressions for axisymmetric solutions. In Sec.~\ref{sec:application} we discuss how the expressions could be used to distinguish between GR and scalar-tensor theory. Finally, in Sec.~\ref{sec:conclusions} we present our conclusions. In the Appendix we give the full expressions for the ISCO radii and the orbital frequencies.

\section{Circular equatorial geodesics in scalar-tensor gravity}
\label{sec:STintro}

One of the most thoroughly studied alternatives to GR is scalar-tensor theory \citep{FujiiMaeda,CapozzielloFaraoni,Damour,Jordan49,fierz56,Jordan59,BransDicke,Dicke62}, which can be described by the following action
\begin{equation}
\label{staction}
S=\int d^4x \sqrt{-g} \left(\Phi R-\frac{\omega(\Phi)}{\Phi} \nabla^\mu \Phi \nabla_\mu \Phi\right)+S_m(g_{\mu\nu},\psi)\,,
\end{equation}
where $g$ is the determinant and $R$ is the 
Ricci scalar of the metric $g_{\mu\nu}$,  
$\nabla_\mu$ denotes the corresponding covariant 
derivative, $S_m$ is the matter Lagrangian, and $\psi$ 
collectively denotes the matter fields.  It is  assumed 
that matter fields couple minimally to
$g_{\mu\nu}$.

In action (\ref{staction}) the scalar field $\Phi$ is nonminimally coupled to gravity and has a noncanonical kinetic term.  This representation of the theory is called the Jordan frame. The conformal transformation $\tilde{g}_{\mu\nu}=16 \pi G\, \Phi
\, g_{\mu\nu}$, together with the scalar field redefinition
\be
\label{scalarredef}
d\phi=\sqrt{\frac{2\omega(\Phi)+3}{4}} \, d\ln\Phi\,,
\ee
will bring action (\ref{staction}) into the following form
\begin{equation}
\label{stactionein}
S=\frac{1}{16\pi G}\int d^4x \sqrt{-\tilde{g}} \left(\tilde{R}-2 \tilde{\nabla}^\mu \phi \tilde{\nabla}_\mu \phi\right)+S_m(g_{\mu\nu},\psi)\,,
\end{equation}
where the matter fields still couple minimally to $g_{\mu\nu}$. This implies that $\phi$ is now coupled to the matter fields, and it is this coupling that encodes any deviation from standard GR with a minimally coupled scalar field. This frame is called the Einstein frame.
The advantage of the Einstein frame is that the field equations outside the matter sources take the form,
\begin{align}
\label{feq1}
\tilde{R}_{ab} &= 2 \partial_a\phi\partial_b\phi,\\
\tilde{g}^{ab}\tilde{\nabla}_a\tilde{\nabla}_b\phi &=0,
\end{align}  
which are essentially GR with a minimally coupled scalar field. 

A spacetime is stationary and axisymmetric if it admits a timelike Killing vector field $\xi^a$ and a spacelike Killing vector field $\eta^a$ that has closed integral curves. The actions of these symmetries should also commute, i.e., $\eta^a\nabla_a\xi^b-\xi^a\nabla_a\eta^b=0$. Furthermore, the condition for the 2-planes that are orthogonal to the two Killing vectors to be integrable is (see \cite{stephani,Wald:1984cw}):
\be \epsilon_{abcd}\eta^a\xi^b\nabla^d\xi^c = 0 = \epsilon_{abcd}\xi^a\eta^b\nabla^d\eta^c .\ee
This condition can be also written in terms of the Ricci tensor in the form:
\be
\label{condition}
\xi^d R_{d[a}\xi_b\eta_{c]} = 0 = \eta^d R_{d[a}\xi_b\eta_{c]}. \ee 
In GR this condition is satisfied in vacuum. As a consequence, the line element of stationary and axisymmetric vacuum spacetimes can take the Weyl-Papapetrou form \citep{papapetrou}, 
\be ds^2=-f(dt-wd\varphi)^2+f^{-1}\left[e^{2\gamma}\left(d\rho^2+dz^2\right)+\rho^2d\varphi^2\right] \label{weyl},\ee  
without loss of generality, where the metric functions depend only on the coordinates $(\rho,z)$. 

In the case of scalar-tensor theories, one can use the vacuum field equations in the Einstein frame, eq.~(\ref{feq1}), in order to show that condition (\ref{condition}) is satisfied.  By virtue of the assumption that the scalar field obeys the symmetries of the metric, i.e., $\xi^a\nabla_a\phi=0$ and $\eta^a\nabla_a\phi=0$, one has that $\xi^a R_{ab}=\eta^a R_{ab}=0$. This implies that the conditions for integrability are satisfied and the line element for a stationary and axisymmetric spacetime in scalar tensor theory can be written in the Weyl-Papapetrou form, eq.~(\ref{weyl}), without any loss of generality. Interestingly enough, as it is shown in \cite{PapSot:STmoments} one can write the field equations for the Weyl-Papapetrou metric in the Einstein frame in the same form as they are in GR. The field equations can, therefore, be written in the form of the equation introduced by \cite{Ernst1}
\be (\mathcal{R}\mathcal{E})\bar{\nabla}^2\mathcal{E}=\bar{\nabla}\mathcal{E}\cdot \bar{\nabla}\mathcal{E},\ee
where $\mathcal{E}=f+i\omega$, $\omega$ is the scalar twist of the timelike Killing vector $\xi^a$ (not to be confused with the $\omega(\Phi)$ in the action of the scalar-tensor theory), and $\bar{\nabla}$ is the gradient in flat cylindrical coordinates.
The Ernst equation is accompanied by two equations for the function $\gamma$ (which we will suppress here for brevity) and one more equation for the scalar field,
\be \bar{\nabla}^2\phi=0, \ee
and this closes the set of equations that characterise a solution of the scalar-tensor theory in the Einstein frame, the solution being a spacetime metric $\tilde{g}_{ab}$ and a scalar field $\phi$.

If we are to study the properties of geodesics that matter follows, we need to go from the metric in the Einstein frame to the metric in the Jordan frame, which is the one with which matter couples minimally. The metric in the Jordan frame can be expressed as the metric in the Einstein frame times a conformal factor that depends on the scalar field on the Einstein frame, $g_{ab}=(16\pi G\Phi)^{-1}\tilde{g}_{ab}=A^2(\phi)\tilde{g}_{ab}$, where the Einstein frame metric is expressed in the Weyl-Papapetrou form, eq.~(\ref{weyl}). The conformal factor $A^2(\phi)$ is a free function of the theory which can be determined by integrating eq.~(\ref{scalarredef}) for a given $\omega(\Phi)$.

Since the scalar field is assumed to respect the symmetries of the metric, the Jordan and the Einstein frame metrics will share the same symmetries. Therefore, in both frames there exist two killing fields which define the symmetries with respect to translations in time and rotations with respect to an axis. One can take advantage of these symmetries to study geodesics in the spacetime in the Jordan frame, as one would in GR.

Symmetry under time translations is associated to an integral of motion, the energy $E$
\be E=-p_a\xi^a=-p_t
   =m\left(-g_{tt}\frac{dt}{d\tau}-g_{t\varphi}\frac{d\varphi}{d\tau}\right),\ee
where $t$ is the coordinate time and $\tau$ is proper time. 
Symmetry under rotations is again associated to an integral of motion, the angular momentum $L$
\be L=p_a\eta^a=p_{\varphi}
   =m\left(g_{t\varphi}\frac{dt}{d\tau}+g_{\varphi\varphi}\frac{d\varphi}{d\tau}\right)\ee 
In addition  to the two previous equations, from the four-momentum of a particle, which is defined as $p^a=mu^a=m\frac{dx^a}{d\tau}$, we have the equation 

\bea -1&=&g_{tt}\left(\frac{dt}{d\tau}\right)^2+2g_{t\varphi}\left(\frac{dt}{d\tau}\right)\left(\frac{d\varphi}{d\tau}\right)
                    +g_{\varphi\varphi}\left(\frac{d\varphi}{d\tau}\right)^2\nn\\
                    &&+ g_{\rho\rho}\left(\frac{d\rho}{d\tau}\right)^2
                    +g_{zz}\left(\frac{dz}{d\tau}\right)^2 \label{4momentum}\eea
If we further define the angular velocity, $\Omega\equiv\frac{d\varphi}{dt}$, for the circular and equatorial orbits, eq.~(\ref{4momentum}) defines the redshift factor between coordinate and proper time,
\be \left(\frac{d\tau}{dt}\right)^2=-g_{tt}-2g_{t\varphi}\Omega -g_{\varphi\varphi}\Omega^2, \label{redshift}\ee
and the energy and the angular momentum for the circular orbits take the form,
\begin{align} \tilde{E}\equiv \frac{E}{\mu}&=\frac{-g_{tt}-g_{t\varphi}\Omega}{\sqrt{-g_{tt}-2g_{t\varphi}\Omega -g_{\varphi\varphi}\Omega^2}},\label{energy}\\
         \tilde{L}\equiv \frac{L}{\mu}&=\frac{g_{t\varphi}+g_{\varphi\varphi}\Omega}{\sqrt{-g_{tt}-2g_{t\varphi}\Omega -g_{\varphi\varphi}\Omega^2}}, \end{align}
where we have introduced the energy and angular momentum per unit mass. 
From the conditions, $\frac{d\rho}{dt}=0,\,\frac{d^2\rho}{dt^2}=0$ and $\frac{dz}{dt}=0$ for circular equatorial orbits, and the equations of motion
obtained assuming the Lagrangian, ${\cal L}=\frac{1}{2}g_{ab}\dot{x}^a\dot{x}^b$, the angular velocity can be calculated to be,
\be
\Omega=\frac{-g_{t\varphi,\rho}+\sqrt{(g_{t\varphi,\rho})^2-g_{tt,\rho}g_{\varphi\varphi,\rho}}}{g_{\varphi\varphi,\rho}}.
\ee
This is the orbital frequency of a particle in a circular orbit on the equatorial plane.            

Equation (\ref{4momentum}) can take a more general form in terms of the constants of motion,
\bea -g_{\rho\rho}\left(\frac{d\rho}{d\tau}\right)^2 -g_{zz}\left(\frac{dz}{d\tau}\right)^2\!\!\!\!\!&=&\!\!\!\!\!
    1-\frac{\tilde{E}^2g_{\varphi\varphi}+2\tilde{E}\tilde{L} g_{t\varphi}+\tilde{L}^2g_{tt}}{(g_{t\varphi})^2-g_{tt}g_{\varphi\varphi}}\nn\\
     \!\!\!\!\!&=&\!\!\!\!\!V_{eff}.\label{eqmotion}\eea
With equation (\ref{eqmotion}) we can study the general properties of the motion of a particle from the properties of the
 effective potential. Furthermore, we can study perturbations around circular equatorial orbits. If we assume small deviations of the form, $\rho=\rho_c+\delta\rho$ and $z=\delta z$, then we obtain the perturbed form of (\ref{eqmotion}),

\bea -g_{\rho\rho}\left(\frac{d(\delta\rho)}{d\tau}\right)^2 -g_{zz}\left(\frac{d(\delta z)}{d\tau}\right)^2\!\!\!\!\!&=&\!\!\!\!\!
       \frac{1}{2}\frac{\p^2 V_{eff}}{\p\rho^2} (\delta \rho)^2 \nn\\
       \!\!\!\!\!&&\!\!\!\!\!+  \frac{1}{2}\frac{\p^2 V_{eff}}{\p z^2} (\delta z)^2. \eea
This equation describes two harmonic oscillators with frequencies,
\begin{align} \bar{\kappa}_{\rho}^2&=\left.\frac{g^{\rho\rho}}{2}\frac{\p^2 V_{eff}}{\p\rho^2}\right|_c\,,\\
                                 \bar{\kappa}_z^2&=\left.\frac{g^{zz}}{2}\frac{\p^2 V_{eff}}{\p z^2}\right|_c\,.\end{align}
The differences of these frequencies (corrected with the redshift factor (\ref{redshift}), i.e., $\kappa_a=(d\tau/dt)\bar{\kappa}_a$) from the orbital frequency,
$\Omega_a=\Omega-\kappa_a$, define the precession frequencies, where the oscillation frequencies $\kappa_a$ are given in terms of the metric functions as,
\bea
\kappa_a^2&=&-\frac{g^{aa}}{2}\left\{(g_{tt}+g_{t\varphi}\Omega)^2\left(\frac{g_{\varphi\varphi}}{A^4(\phi)\rho^2}\right)_{,aa}\right.\nn\\
              &&-2(g_{tt}+g_{t\varphi}\Omega)(g_{t\varphi}+g_{\varphi\varphi}\Omega)\left(\frac{g_{t\varphi}}{A^4(\phi)\rho^2}\right)_{,aa}\\
              &&\left.+(g_{t\varphi}+ g_{\varphi\varphi}\Omega)^2\left(\frac{g_{tt}}{A^4(\phi)\rho^2}\right)_{,aa}\right\}\Bigg |_{z=0}\,,\nn
\eea        
where the index $a$ takes either the value $\rho$ or $z$ to express the frequency of the radial or the vertical perturbation
respectively and the expressions are evaluated on the equatorial plane $z=0$. The position where $\kappa_{\rho}^2$ becomes zero is the location of the ISCO. 
These equations will be used to calculate $R_{\textrm{ISCO}}$, the various frequencies $\Omega_{\rho}$, $\Omega_z$ and $\Omega$, and the quantity $\Delta\tilde{E}\equiv-\frac{d\tilde{E}}{d\ln\Omega}$ in scalar-tensor theory.

There is one final issue that needs to be discussed. Properties associated with geodesics of a metric will depend only on the metric in question and not on the corresponding scalar field. Here we are considering geodesics of the Jordan frame metric, as these are the ones followed by test particles. Their properties will not depend directly on $\Phi$ (obviously  $\Phi$ does implicitly affect the geodesics by acting as a source in the modified Einstein equations). One can be tricked into thinking that the scalar moments will therefore not appear in the expression for the various observables, as the metric is fully determined by its own moments. However, the multipole moments in \cite{PapSot:STmoments} are actually defined in the Einstein frame  and $(16\pi G\,\Phi)^{-1}=A^2(\phi)$ is the conformal factor that relates the Jordan and the Einstein metrics. So, the observables will explicitly depend on both the scalar and metric multipole moments. 

The hidden presence of the free function $A(\phi)$ of the conformal factor also highlights the need to select a specific function $A(\phi)$ in order to pin down the theory in consideration [recall that 
$A(\phi)$ is related to  $\omega(\phi)$ through eq.~(\ref{scalarredef})]. We would like to avoid doing so in order to keep our results as general as possible, so we will adopt the following expansion of the conformal factor around the asymptotic value of scalar field, $\phi_{\infty}=\phi_0$,
\begin{align}
 A(\phi)&= A(\phi_0)+A'(\phi_0)(\phi-\phi_0)+A''(\phi_0)(\phi-\phi_0)^2\nn\\
                &+A^{(3)}(\phi_0)(\phi-\phi_0)^3+\ldots \nn\\
                &=A(\phi_0)\left(1+\frac{A'(\phi_0)}{A(\phi_0)}(\phi-\phi_0)+\frac{A''(\phi_0)}{A(\phi_0)}(\phi-\phi_0)^2\right.\nn\\
                &\left.+\frac{A^{(3)}(\phi_0)}{A(\phi_0)}(\phi-\phi_0)^3+\dots\right).
                \end{align}
Here, we can define in the usual way (see for example \cite{Damour,Damour96PRD}) the parameters $\alpha_0=d\ln A/d\phi |_{\infty}$, $\beta_0=d\alpha/d\phi |_{\infty}$, $\gamma_0=d\beta/d\phi |_{\infty}$, and so on. The derivatives of $A(\phi)$ at infinity can then be expressed in terms of the parameters $\alpha,\beta,\gamma$, etc. (for example, $A''(\phi_0)=A(\phi_0)(\beta+\alpha^2)$).  

With this parameterisation of the conformal factor any observable associated to geodesics in the Jordan frame can be expressed in terms of the moments and the coefficients of the expansion, i.e., $A(\phi_0)$, $\alpha,\beta,\gamma$, etc. Note that the asymptotic value of the Jordan frame scalar $\Phi_\infty$ is related to the asymptotic value of the Einstein frame scalar $\phi_0$ via $(16\pi G\,\Phi_\infty)^{-1}=A^2(\phi_0)$. In what follows, we will set $A(\phi_0)=1$ for simplicity. This amounts to a Weyl rescaling of the Jordan frame metric and it should be taken into consideration when one interprets the final expressions for the observables. In particular, according to their definitions $\Omega$, $\Omega_\rho$, and $\Omega_z$ are insensitive to such a rescaling, whereas $\Delta E/\mu$ picks up an $A(\phi_0)^{-1}$ factor if $\mu$ is held fixed. $\mu$  actually corresponds to some particle mass, $\mu^2=-p_ap^a$, so in reality this mass would be also rescaled by a factor $A(\phi_0)^{-1}$ and the ratio $\Delta E/\mu$ would be insensitive to constant Weyl rescalings. Caution should also be taken when calculating $\alpha,\beta,\gamma$, etc. for a given theory, as their values are affected by such rescalings.

Finally, we should stressed that the multipole moments for the scalar field are defined up to a constant shift of the Einstein scalar, {\em i.e.}~they actually characterise the quantity $\phi-\phi_0$. When calculating the moments one can set $\phi_0=0$ without loss of generality and this is the approach we will follow in the rest of the paper. As is obvious from the above though, the actual value of $\phi_0$ does affect the observables through $A(\phi_0)$ and the higher order coefficients of the expansion for $A(\phi)$ and this will be taken consistently into account.

\section{Observables and moments in spherically symmetric solutions}
\label{sec:spherical}

As a warm-up calculation, we will start with the static, spherically symmetric solution in scalar-tensor theory. 
The metric, expressed in the Einstein frame, is 
\begin{align} ds^2=&-\left(1-\frac{2l}{r}\right)^{\frac{m}{l}} dt^2+\left(1-\frac{2l}{r}\right)^{-\frac{m}{l}} dr^2\nn\\
&+\left(1-\frac{2l}{r}\right)^{1-\frac{m}{l}} r^2(d\theta^2+\sin^2\theta d\varphi^2). \end{align}
In these coordinates, the corresponding scalar field is
\be
\phi=\frac{w_A}{2l}\log\left(1-\frac{2l}{r}\right)\,,
\ee 
and $m^2+w_A^2=l^2$, {\em i.e.}~this is actually a 2-parameter family of solutions.
This is the static, spherically symmetric solution in scalar-tensor theory that is given in \cite{Just1959ZNat, Damour}.\footnote{In the notation of \cite{Damour} $2l$ corresponds to $a$,  $2m$, corresponds  $b$, and  the scalar charge $w_A$ corresponds to $d$. The bond $m^2+w_A^2=l^2$  corresponds to  $a^2-b^2=4d^2$. The comparison also reveals that $m$ is the Einstein frame mass.} 

The metric written in Weyl-Papapetrou coordinates  takes the form
\begin{align} 
\label{wpmetric}
ds^2=&-\left(\frac{R_- +R_+ -2l}{R_-+R_++2l}\right)^{\frac{m}{l}}dt^2+\left(\frac{R_- +R_+ -2l}{R_-+R_++2l}\right)^{-\frac{m}{l}}\nn\\
            &\times\left[\left(\frac{1}{2} +\frac{z^2+\rho^2-l^2}{2R_+ R_-}\right)(d\rho^2+dz^2)+\rho^2 d\varphi^2\right],\quad
\end{align} 
where
\begin{align} 
R_+\equiv \sqrt{(l-z)^2+\rho^2}\,,\\
R_-\equiv\sqrt{(l+z)^2+\rho^2}\,.
\end{align}
The scalar profile in the same coordinates is
\begin{equation}
\label{scalaransatz}
\phi=\frac{w_A}{2l} \log \left(\frac{\sqrt{(l+z)^2+\rho ^2}-l-z}{\sqrt{(l-z)^2+\rho^2}+l-z}\right)\, ,
\end{equation}   
where $w_A$ turns out to be the scalar charge. 

In order to calculate the orbital parameters that one would measure for test particles moving along geodesics in this spacetime, one needs to use the metric in the Jordan frame, which is conformally related to the metric in the Einstein frame through the relation $g_{ab}=A^2(\phi)\tilde{g}_{ab}$. In this expression then, one can substitute $A(\phi)$ with its Taylor expansion in powers of $(\phi-\phi_0)$, which can then be replaced by eq.~\eqref{scalaransatz}.

Following \cite{Ryan95} we would like to express the various observables as series expansions in the orbital frequency $\Omega$. The coefficients of such an expansion will end up depending on the multipole moments and one could in principle then determine the latter order-by-order from observations. The orbital frequency of circular orbits on the equatorial plane ($z=0$) can be shown to have the following expansion in inverse powers of $\rho$,
\be \Omega=\left(\frac{M+w_A \alpha_0}{\rho^3}\right)^{1/2}\left(1+\textrm{series in } \rho^{-1/2}\right). \ee
We can see from the expansion that we can redefine the mass as
\be
\tilde{M}=M+w_A \alpha_0\,.
\ee
It is preferable to work in terms of the dimensionless quantities  $U=(\tilde{M}\Omega)^{1/3}$ and $x=(\tilde{M}/\rho)^{1/2}$. $U$ can be expressed as a series expansion in $x$ as 

\begin{align}
 U=& x+\frac{1}{6} \left[\frac{w_A \left(2 \alpha_0  \tilde{M}-\beta_0 
   w_A\right)}{\tilde{M}^2}-3\right]x^3\nn\\
&+\frac{x^5}{72 \tilde{M}^4}\left[2 w^3 \tilde{M} (3 \gamma_0 -2 \alpha_0  \beta_0 )\right.\nn\\&+2 w^2
   \tilde{M}^2 \left(11 \alpha_0 ^2+9 \beta_0 -3\right)\nn\\
 &\left.-24 \alpha_0  w \tilde{M}^3+9 \tilde{M}^4-5 \beta_0 ^2
   w^4\right]
   +O\left(x^7\right)
   \end{align}
This series expansion can be inverted and $x$ can be expressed as an expansion in $U$. Observables can then be expressed as power series in $U$. Note that $U$ is preferable to $\tilde{M}\Omega$ because the expansion will contain integer powers of $U$, as opposed to fractional powers of $\tilde{M}\Omega$.

With the strategy laid out, we can proceed to deriving the actual expressions for the various observables, starting with  $\Delta E/\mu$. The energy per unit mass of circular orbits is given by eq.~(\ref{energy}).
%
Once this is expressed as a series expansion in $U$, the energy per logarithmic frequency interval is given as 
\be \frac{\Delta E}{\mu}=-\frac{U}{3}\frac{dE}{dU}, \label{DEeq1}\ee
which as a series has the following expansion,
\begin{align}
 \frac{\Delta E}{\mu}=& \frac{U^2}{3}+\left[\frac{2 \beta_0  w_A^2}{9 \bar{M}^2}+\frac{8 \alpha_0  w_A}{9
   \bar{M}}-\frac{1}{2}\right]U^4 
+\Bigg[8 \beta_0 ^2
   w_A^4 \nn\\
&   -8 \bar{M} w_A^3 (4 \alpha_0  \beta_0 +\gamma_0 )+4 w_A^2 \bar{M}^2
   \left(2-38 \alpha_0 ^2+3 \beta_0 \right) \nn\\
& +224 \alpha_0  w_A \bar{M}^3-81 \bar{M}^4\Bigg]\frac{U^6 }{24 \bar{M}^4}
   +O\left(U^8\right).   
\end{align}
If one sets the scalar field charge to zero, one recovers the Schwarzschild expansion in GR. It is already evident in this expression that one cannot do away with the effects of the corrections coming from the scalar field by a redefinition of the mass.

One can proceed to calculate the expansions for the two epicyclic frequencies, $\Omega_{\rho}$ and $\Omega_z$. The periastron precession will give the expansion,
\begin{align}
 \frac{\Omega_{\rho}}{\Omega}= &
 \left[3-\frac{w_A \left(8 \alpha_0  \tilde{M}+\beta_0  w_A\right)}{2 \tilde{M}^2}\right]U^2\nn\\
&\!+ \Bigg[4
   \tilde{M} \Big(2 w_A^2 \tilde{M} \left(17 \alpha_0 ^2+3 \beta_0 -3\right)-60 \alpha_0  w_A \tilde{M}^2\nn\\
&\! +27 \tilde{M}^3 +3 \gamma_0  w_A^3\Big)-13 \beta_0 ^2 w_A^4\Bigg] \frac{U^4 }{24 \tilde{M}^4}\nn\\
&\!+\Bigg[ 4 \beta_0  w_A^5 \tilde{M} (2 \alpha_0  \beta_0 +11 \gamma_0 )-6 w_A^4 \tilde{M}^2 \big(20 \alpha_0 ^2 \beta_0 +6 \alpha_0  \gamma_0 \nn\\
& \!  +\beta_0 ^2+4 \beta_0 +2 \delta_0 \big)+8 w_A^3 \tilde{M}^3 \Big(-112 \alpha_0 ^3+2 \alpha_0  (\beta_0 +8)\nn\\
&\!+3 \gamma_0  \Big)+12 w_A^2 \tilde{M}^4 \left(196 \alpha_0 ^2+7 \beta_0 -12\right)-2112 \alpha_0  w_A \tilde{M}^ 5 \nn\\
&\!+648 \tilde{M}^6-35  \beta_0 ^3 w_A^6   \Bigg]\frac{U^6}{48 \tilde{M}^6} +O\left(U^7\right)
\end{align}
%
%
Again, this expression reduces to the Schwarzschild one if we set the scalar charge equal to zero. 

As is the case in GR, the spherically symmetric solution in scalar-tensor leads to zero nodal precession, since $\kappa_z=\Omega$  and, therefore, $\Omega_z/\Omega=0$.\footnote{In the work by \cite{DonevaSTprecess2014}, one can find analytic expressions for the orbital frequency and the radial oscillation frequency of the spherically symmetric spacetime in scalar-tensor theory, for a specific choice of the conformal factor.} 

What remains to be calculated, is the location of the ISCO. One way of identifying the location of the ISCO is as the radius at which the radial oscillation frequency of a slightly eccentric orbit becomes zero. Therefore one needs to calculate the roots of the square of the frequency $\kappa_{\rho}^2$. Even in the simple case of spherical symmetry, the evaluation of the roots is not easy and there is additionally the problem of having an unspecified function, which is the conformal factor for the transformation to the Jordan frame, $A(\phi)$. This problem can be solved using a perturbative approach as long as the compact object is not considered to be too scalarized. In that case, one could treat the scalar charge (actually the scalar charge over the mass) as a perturbative quantity of order $\varepsilon$, while at the same time assume that the sought after root has the GR value plus corrections, i.e., $ \rho_{\textrm{{\tiny ISCO}}}=2\sqrt{6}M(1+c_1\varepsilon+c_2\varepsilon^2+\ldots)$. This expression for the radius can be substituted in the function $\kappa_{\rho}^2$, which can then be expanded as a series in $\epsilon$. The result will be a system of equations that will have to be solved in terms of the coefficients $c_i$ so that the expansion gives zero at all orders. This calculation is analogous to the calculation presented by \cite{SSisco}. The resulting expression up to fourth order for the ISCO is,

\begin{align}
\rho_{\textrm{{\tiny ISCO}}}=& 2 \sqrt{6} M \Big(1-\frac{5 \alpha_0  w_A}{16 M}+\left(0.0371084 \alpha_0 ^2-0.092897 \beta_0 \right.\nn\\
 &\left. -0.072918\right)\frac{w_A^2 }{M^2}+\Big[0.009646 \alpha_0 ^3+\alpha_0  (0.197845 \beta_0 \nn\\
 &  -0.0201845)+0.025255 \gamma_0 \Big]\frac{w_A^3 }{M^3}+\Big[0.000736 \alpha_0 ^4\nn\\
 & +\alpha_0 ^2 (-0.185728 \beta_0
   -0.0000175)-0.0290932 \alpha_0  \gamma_0 \nn\\
  &  +(-0.0225927 \beta_0 -0.001873) \beta_0 -0.002777 \delta_0 \nn\\
 &   -0.0046151\Big]\frac{w_A^4 }{M^4}+\ldots \Big)
\end{align}
which in terms of circumferential radius can be written as,
\begin{align}
 R_{\textrm{{\tiny ISCO}}}=& 6 M\Big(1-\frac{0.452733 \alpha_0  w_A}{M}+\left(0.0399826 \alpha_0 ^2\right.\nn\\
&\left.-0.053767 \beta_0
   -0.059701\right)\frac{w_A^2 }{M^2}+\left(0.0115299 \alpha_0 ^3\right.\nn\\
   &\left.+\alpha_0  (0.159059 \beta_0
   -0.017441)+0.018815 \gamma_0 \right)\frac{w_A^3 }{M^3}\nn\\
   &+\left[\alpha_0 ^2 (0.0001688\, -0.143397 \beta_0 )+0.001467 \alpha_0 ^4\right.\nn\\
   &-0.0232982
   \alpha_0  \gamma_0 -(0.0154855 \beta_0 +0.000099) \beta_0 \nn\\
   &\left.-0.0021512 \delta_0
   -0.003793\right]\frac{w_A^4 }{M^4}+\ldots \Big)
\end{align}
Finally, one last useful quantity at the ISCO that we could evaluate in terms of the parameters of the spacetime is the orbital frequency, 
\begin{align}
\Omega_{\textrm{{\tiny ISCO}}}=&\frac{1}{6 \sqrt{6} M}\Big(1+\frac{0.75 \alpha_0  w_A}{M}+ \left(1.21083\times10^{-6} \alpha_0
   ^2\right.\nn\\
   &\left.+0.0354517 \beta_0 +0.0777343\right)\frac{w_A^2}{M^2}+\left(\alpha_0  [0.089554 \right.\nn\\
   &\left. -0.117187 \beta_0 ]+0.007811 \alpha_0 ^3-0.022599 \gamma_0
   \right)\frac{w_A^3 }{M^3}\nn\\
   &+\left(-0.0424804 \alpha_0 ^4+\alpha_0 ^2 (0.009887 \beta_0 +0.023437)\right.\nn\\
   &+0.009229
   \alpha_0  \gamma_0 +(0.0178996 \beta_0 +0.006044) \beta_0 \nn\\
   &\left.+0.002812 \delta_0
   +0.00983\right)\frac{w_A^4 }{M^4}+\ldots \Big)
\end{align}

One could ask whether the assumption of a weakly scalarized compact object is a reasonable one. The current constrains on the values of the parameters $\alpha_0, \beta_0$ for scalar-tensor theories are such that do not allow for very large scalarization for non-rotating compact objects. Therefore, treating the scalar field as a perturbative parameter in the non-rotating case is a reasonable assumption [for example see \cite{will-living,Freire2012STconstrains,Doneva2013PhRvD,PaniBerti2014PhRvD,BertietalNS2015}].

\section{Observables and moments in axisymmetric solutions}
\label{sec:generalST}

We will now derive the general expressions that relate the orbital frequency to the precession frequencies and the energy change per logarithmic frequency interval for a stationary and axisymmetric spacetime in scalar-tensor theory of gravity. We will also present expressions for the location of the ISCO and the orbital frequency at the ISCO which will hold under certain reasonable assumptions regarding the behaviour of the higher order moments of both the spacetime and the scalar field.

\subsection{Ernst potential, scalar field and moments}

As we have briefly discussed in Section \ref{sec:STintro}, in the Einstein frame the stationary and axisymmetric solutions of scalar-tensor theory for a massless scalar field, can be described in terms of an Ernst potential and a scalar field. In particular, as it is described by \cite{PapSot:STmoments}, the Ernst potential and the scalar field of the solution are fully determined by their values along the axis of symmetry, which can in turn be given in terms of the multipole moments of both the scalar field and the metric. We remind that in the calculation of the moments the auxiliary potential $\xi=(1-\mathcal{E})/(1+\mathcal{E})$ is also used.

We define the coordinates at infinity 
$\bar{\rho}=\rho/(\rho^2+z^2)$,   $\bar{z}=z/(\rho^2+z^2)$
where $\bar{r}^2=\bar{\rho}^2+\bar{z}^2$. The potential $\tilde{\xi}=(1/\bar{r})\xi$ can be expressed as a series expansion around infinity of the form,
\be \tilde{\xi}=\sum_{i,j=0}^{\infty} a_{ij}\bar{\rho}^i\bar{z}^j, \ee
where the coefficients $a_{ij}$ can be expressed with respect to the coefficients $a_{0j}=m_j$ of the expansion of $\tilde{\xi}$ along the axis of symmetry, $\tilde{\xi}(\bar{\rho}=0)=\sum_{j=0}^{\infty} m_j\bar{z}^j$. The coefficients $m_j$ are related to the multipole moments of the spacetime and can be found by solving the expressions given by \cite{PapSot:STmoments}. In a similar way, one can express $\tilde{\phi}=(1/\bar{r})\phi$ as a series expansion at infinity of the form,
\be \tilde{\phi}=\sum_{i,j=0}^{\infty} b_{ij}\bar{\rho}^i\bar{z}^j,\ee
and one can see that the coefficients $b_{ij}$ can be calculated in terms of the coefficients $b_{0j}=w_j$ of the expansion of $\tilde{\phi}$ along the symmetry axis, $\tilde{\phi}(\bar{\rho}=0)=\sum_{j=0}^{\infty} w_j\bar{z}^j$. The $w_j$ coefficients are related to the scalar field moments and can be also calculated from the expressions given by \cite{PapSot:STmoments}. 

Once we have the Ernst potential and the scalar field,

\be \mathcal{E}=\frac{r-\tilde{\xi}(\rho,z)}{r+\tilde{\xi}(\rho,z)}, \;\; \phi=\frac{\tilde{\phi}(\rho,z)}{r},\ee
where $r=\sqrt{\rho^2+z^2}$, we can proceed to calculate the metric functions. 

The function $f(\rho,z)$ of the metric \eqref{weyl} is the real part of the Ernst potential $f=\Re(\mathcal{E})$, while the imaginary part of the Ernst potential is $\omega=\Im(\mathcal{E})$. This $\omega(\rho,z)$ function is related to the metric function $w(\rho,z)$, in \eqref{weyl}, by the identity

\be f^{-2}\bar{\nabla}\omega=-\rho^{-1}\hat{n}  \times\bar{\nabla}w, \ee
where $\bar{\nabla}$ is the gradient in the cylindrical flat coordinates $(\rho,z,\varphi)$ and $\hat{n}$ is a unit vector in the azimuthal direction. This equation can be integrated to give the metric function $w(\rho,z)$ in terms of the moments (for example see the discussion by \cite{Ryan95}). The only metric function that remains to be determined is the function $\gamma(\rho,z)$ and it can be evaluated by integrating eqs.~(47)-(48) in \cite{PapSot:STmoments}, in the same way as one would do in GR (as before, see the discussion by \cite{Ryan95}).

This calculation will give us the Einstein frame metric $\tilde{g}_{ab}$ in terms of the moments. The Jordan frame metric, which is the metric to which particles and photons couple,  will then be given as $g_{ab}=A^2(\phi)\tilde{g}_{ab}$. This metric is now expressed in terms of the spacetime multipole moments and the moments of the scalar field, and is the metric that we will use to calculate all the quantities that are related to the geodesics.   

We should note here, that we assume that particles and photons follow the geodesics of the Jordan frame metric and that there are no interactions of the particles and photons with the scalar field. If we were to assume such interactions, then the orbits would deviate from being geodesic.

\subsection{Energy and frequencies of equatorial orbits}

In what follows, we will adopt the following convention for the multipole moments of the spacetime and the moments of the scalar field in connection to the moments defined by \cite{PapSot:STmoments}: the mass moments will be $M_a\equiv \Re(P_a^g)$, where in particular we will have for the mass $M_0\equiv \Re(P_0^g)=m_0\equiv M$, the angular momentum moments will be $J_a\equiv \Im(P_a^g)$, where the angular momentum will be $J_1\equiv \Im(P_1^g)=\Im(m_1)$, and finally for the scalar field moments we will have $W_a\equiv P_a^{\phi}$, where the scalar monopole will specifically be $W_0\equiv P_0^{\phi}=w_0$. 

We proceed as we did in the spherical case by evaluating the orbital frequency of circular equatorial orbits. In this case the expansion of the orbital frequency in inverse powers of $\rho$ will be 
\be \Omega=\left(\frac{M-W_0 \alpha_0}{\rho^3}\right)^{1/2}\left(1+\textrm{series in } \rho^{-1/2}\right). \ee
Here we redefine the mass as $\bar{M}=M-W_0 \alpha_0$, noting that there is a difference in the sign convention for the scalar field that results in having essentially $W_0=-w_A$. Again, we expand $U=(\bar{M}\Omega)^{1/3}$ as a series in powers of $x=(\bar{M}/\rho)^{1/2}$ and invert the expansion so that we may express $x$ as a series in powers of $U$. From eqs. \eqref{energy} and \eqref{DEeq1} we finally get to the expansion of the energy change per logarithmic frequency interval change in terms of $U$,

\begin{align} \frac{\Delta E}{\mu}=&
\frac{U^2}{3}+\left(\frac{2 \beta_0  W_0^2}{9 \bar{M}^2}-\frac{8 \alpha_0  W_0}{9
   \bar{M}}-\frac{1}{2}\right)U^4 +\frac{20 J_1 U^5}{9 \bar{M}^2}\nn\\
&  +\Big(8 \bar{M} \left[3
   M_2+W_0^3 (4 \alpha_0  \beta_0 +\gamma_0 )-3 \alpha_0  W_2\right]\nn\\
&+4 W_0^2 \bar{M}^2
   \left(-38 \alpha_0 ^2+3 \beta_0 +2\right)-224 \alpha_0  W_0 \bar{M}^3\nn\\
& -81 \bar{M}^4+8 \beta_0 ^2
   W_0^4\Big)\frac{U^6 }{24 \bar{M}^4} \nn\\
  & +\frac{28 J_1 \left(10 \alpha_0  W_0 \bar{M}+9 \bar{M}^2+2
   \beta_0  W_0^2\right)}{27 \bar{M}^4}U^7 +O\left(U^8\right).
\end{align}
This expression reduces to the one corresponding to the non-rotating case of the previous section if we set to zero the higher moments and choose the scalar monopole to be $W_0=-w_A$ as we have already mentioned.

In the same way, we can proceed to calculate the expansions related to the two precession frequencies, i.e., $\Omega_{\rho}/\Omega$ and $\Omega_z/\Omega$. These expressions are
\begin{align}
 \frac{\Omega_{\rho}}{\Omega}=&
\left(3-\frac{W_0 \left(\beta_0  W_0-8 \alpha_0  \bar{M}\right)}{2 \bar{M}^2}\right)U^2 -\frac{4 J_1 U^3}{\bar{M}^2}\nn\\
&\!\! +\Big(-12 \bar{M} \left(3 M_2+\gamma_0  W_0^3-3 \alpha_0 
   W_2\right)+240 \alpha_0  W_0
   \bar{M}^3\nn\\
&\!\!+8 W_0^2 \bar{M}^2 \left(17 \alpha_0 ^2+3 \beta_0 -3\right)+108 \bar{M}^4-13 \beta_0 ^2 W_0^4\Big)\frac{U^4 }{24 \bar{M}^4} \nn\\
   &\!\!
 -\frac{2 \left[J_1 \left(11
   \alpha_0  W_0 \bar{M}+15 \bar{M}^2+5 \beta_0  W_0^2\right)\right]}{3 \bar{M}^4} U^5 \nn\\
&\!\! +\Big(-6
   \bar{M}^2 \left[16 J_1^2+W_0 \left(24 \alpha_0  M_2+20 \alpha_0 ^2 \beta_0  W_0^3\right.\right.\nn\\
   &\!\! \left.\left.+(6 \alpha_0  \gamma_0+\beta_0 ^2 +4 \beta_0 +2 \delta_0  )W_0^3-24 \alpha_0 ^2 W_2-24
   \beta_0 W_2\right)\right]\nn\\
 &\!\!-8 \bar{M}^3 \left[63 M_2+W_0^3 \left(-112 \alpha_0 ^3+2 \alpha_0 
   (\beta_0 +8)+3 \gamma_0 \right)\right.\nn\\
   &\!\!\left. -9 \alpha_0  W_2\right]-4 \beta _0 W_0^2 \bar{M} \left[33
   M_2+W_0^3 (2 \alpha_0  \beta_0 +11 \gamma_0 )\right.\nn\\
   &\!\!\left.-33 \alpha_0 W_2\right]+12 W_0^2 \bar{M}^4
   \left(196 \alpha_0 ^2+7 \beta_0 -12\right)\nn\\
   &\!\!+2112 \alpha_0 W_0 \bar{M}^5+648 \bar{M}^6-35 \beta_0 ^3
   W_0^6\Big)\frac{U^6 }{48 \bar{M}^6}\nn\\
   &\!\!-\Big(30 J_1 \bar{M} \left[15 M_2+W_0^3 (\alpha_0
    \beta_0 +5 \gamma_0 )-15 \alpha_0  W_2\right]\nn\\
    &\!\!+\bar{M}^2 \Big[2 J_1 W_0 \left(5 W_0
   \left[376 \alpha_0 ^2+33 \beta_0 +39\right]-243\right)\nn\\
   &\!\!-810 J_3\Big]+7980 \alpha_0  J_1 W_0
   \bar{M}^3+4320 J_1 \bar{M}^4\nn\\
   &\!\!+455 \beta_0 ^2 J_1 W_0^4\Big)\frac{U^7 }{90 \bar{M}^6}  +O\left(U^8\right)
\end{align}
and
\begin{align} \frac{\Omega_z}{\Omega}=&
\frac{2 J_1 U^3}{\bar{M}^2}\!+\!\frac{3 U^4 (M_2-\alpha_0  W_2)}{2 \bar{M}^3}\!+\!\frac{2 J_1
  W_0 \left(\beta_0  W_0-\alpha_0  \bar{M}\right) U^5 }{\bar{M}^4}\nn\\
   &+\Big(\bar{M} \Big\{6
   M_2 \bar{M}+14 J_1^2+W_0 \Big[(2 \alpha_0 ^2 -3 \beta_0) 
   W_2  \nn\\
&-2 \alpha_0  M_2\Big]\Big\}+5 \beta_0  W_0^2 (M_2-\alpha_0  W_2)\Big)\frac{U^6 }{2
   \bar{M}^5}\nn\\
&   +\Big(\bar{M} \Big\{5 J_1 \Big[33 M_2+W_0^3 (3 \gamma_0 -4 \alpha_0  \beta_0 )-33 \alpha_0 
   W_2\Big]\nn\\
   &-2 \bar{M} \Big[30 \alpha_0  J_1 W_0 \bar{M}+J_1 W_0 \left(5 [5
   \alpha_0 ^2-6] W_0+27\right)\nn\\
   &+45 J_3\Big]\Big\}+40 \beta_0 ^2 J_1
   W_0^4\Big)\frac{U^7
   }{15 \bar{M}^6}  +O\left(U^8\right)
\end{align}
Again we can see here that if we take the angular momentum and all the higher order multipole moments to zero, then the nodal precession expression, $\Omega_z/\Omega$, reduces to zero, while the periastron precession expression, $\Omega_{\rho}/\Omega$, reduces to the expression for the non-rotating case (taking into account the sign reversal in the scalar monopole). 


\subsection{Location of the last stable circular orbit}
\label{sec:isco}

We now turn to the estimation of the location of the ISCO as the locus of points where the radial oscillation frequency vanishes. The situation in this case is a little more complicated than what we saw in the spherically symmetric case. The perturbative treatment that we followed in the spherically symmetric case used the scalar charge over the mass as an expansion parameter, $\varepsilon$, and made the ansatz $ \rho_{\textrm{{\tiny ISCO}}}=2\sqrt{6}M(1+c_1\varepsilon+c_2\varepsilon^2+\ldots)$ for the location of the ISCO. With this ansatz one can solve equation $\kappa_{\rho}^2=0$ order by order in $\varepsilon$ for the various $c_i$ coefficients. Once one abandons the assumption of spherical symmetry, one needs to also account for the rotation. The most accurate way to proceed would be to consider two different expansion parameters, one associated to rotation, where an appropriate order in $\varepsilon_1$ would be introduced to all higher order moments, and one to scalarization, where an appropriate order in $\varepsilon_2$ would be introduced to all scalar moments. 

This would obviously complicate things and would not allow for the same treatment as in GR or as we had in the spherical case. Therefore, we will prefer to stick to using only one expansion parameter and attempt to assign an appropriate order in this parameter $\varepsilon$ to all higher order moment, both those of the metric and those of the scalar field. Clearly, this will limit significantly the applicability of our treatment, but it will provide some insight into how the scalar moments affect the location of the ISCO. Note that, even though a detailed study of how the various moments behave for scalarized stars is still pending, the preliminary results that already exist in the literature, e.g. the work by \cite{Doneva2013PhRvD}, can help us make an educated guess when assigning an order in $\varepsilon$ to the various moments. 

Starting with the behaviour of the spacetime moments, we will treat the rotation as a perturbation and we will introduce an expansion parameter associated to the angular momentum of the star. Therefore, we will assume for the spacetime moments that the angular momentum $J_1$ will be of order $\varepsilon$, the quadrupole moment $M_2$ will be of order $\varepsilon^2$, the spin octupole $J_3$ will be of order $\varepsilon^3$, and the mass hexadecapole $M_4$ will be of order $\varepsilon^4$. This assumption is justified by the fact that, as it has been shown by \cite{poisson,pappas-apostolatos,YagietalM4} for compact objects, the moments higher than the angular momentum scale as $j^n$, where $j=J_1/M^2$ is the spin parameter. Therefore, if we assume that our perturbative parameter is the rotation, then the rest of the moments should scale as the corresponding powers of $j$. Clearly, this behaviour is what we see in GR and it is not necessary that it will persist in scalar-tensor theory. However, in absence of a better guess it should serve as a good first approximation.

The situation is more complicated for the multipole moments of the scalar field. As we argued previously, we would like to keep one expansion parameter instead of introducing two, but we have essentially two types of effects, the scalarization and the deformation due to rotation. With respect to rotation, \cite{Doneva2013PhRvD} have shown that it introduces some ellipticity in the profile of the scalar field with respect to the spherical profile of the non-rotating case. Therefore, one expects that the higher order moments of the scalar field will receive a contribution that scales with the angular momentum in the same way as the corresponding mass moments of the spacetime, i.e., the contribution to the scalar quadrupole $W_2$ will be of order $\varepsilon^2$ and the contribution to the scalar hexadecapole $W_4$ will be of order $\varepsilon^4$. If we were to keep the scalar charge as an independent expansion parameter, $\varepsilon_1$, then a reasonable guess for the behaviour of the  scalar moments would be the following: the scalar charge would be $W_0\sim\varepsilon_1$, the scalar quadrupole would be $W_2\sim\varepsilon_1\varepsilon^2$, and the scalar hexadecapole would be $W_4\sim\varepsilon_1\varepsilon^4$. 

In order to maintain one expansion parameter one could choose to look at the regime where $\varepsilon\sim \varepsilon_1$. This is the regime where the influence of the scalar charge is comparable to that of the rotation. In this case, we should assign to the scalar monopole an order of $W_0\sim\varepsilon$, while the higher order moments will scale as, $W_2\sim\varepsilon^3$ for the scalar quadrupole and $W_4\sim\varepsilon^5$ for the scalar hexadecapole. This scaling of the scalar moments might seem reasonable, but it is at best just an educated guess. One could imagine a situation where at a more rapid rotation rate the higher scalar moments scale with rotation at a weaker power of $\varepsilon$. Therefore an alternative scaling that one could consider would be to assign to the monopole an order of $W_0\sim\varepsilon$, but for the higher order moments to assume that they scale at one order lower than in the previous case, having therefore a scalar quadrupole that behaves as $W_2\sim\varepsilon^2$ and a scalar hexadecapole that behaves as $W_4\sim\varepsilon^4$. Of course this is an ad hoc choice, but we will use it as an alternative example for illustrative purposes.


Using the first of the two different scalings for the multipole moments and the usual ansatz for the position of the ISCO, i.e., $ \rho_{\textrm{{\tiny ISCO}}}=2\sqrt{6}M(1+c_1\varepsilon+c_2\varepsilon^2+\ldots)$, the resulting expression up to fourth order for the ISCO is,\footnote{Here we present the expressions up to terms of order $M^{-4}$. The full expressions are given in the Appendix.}
 
\begin{align} \rho_{\textrm{{\tiny ISCO}}}=& 2 \sqrt{6} M \Big(1+\frac{5  \alpha_0  W_0}{16 M}-\frac{5 J_1}{3 \sqrt{6} M^2}+\frac{0.0372978 \alpha_0 ^2 W_0^2}{M^2}\nn\\
   -&\frac{0.092923 \beta_0  W_0^2}{M^2}-\frac{0.072877 W_0^2}{M^2}-\frac{0.21977 M_2}{M^3}\nn\\
   -&\frac{0.009463 \alpha_0 ^3 W_0^3}{M^3}-\frac{0.197912 \alpha_0  \beta_0  W_0^3}{M^3}+\frac{0.019801 \alpha_0  W_0^3}{M^3}\nn\\
   -&\frac{0.025259 \gamma_0    W_0^3}{M^3}+\frac{0.073633 \alpha_0  W_2}{M^3}-\frac{0.162089 \alpha_0    J_1 W_0}{M^3}\nn\\
   +&\frac{0.068131 J_3}{M^4}-\frac{0.289247  J_1^2}{M^4}-\frac{0.013148 \alpha_0  M_2  W_0}{M^4}\nn\\
   -&\frac{0.181816 \alpha_0 ^2 J_1 W_0^2}{M^4}-\frac{0.037135 \beta_0   J_1 W_0^2}{M^4}-\frac{0.0678245 J_1 W_0^2}{M^4}\nn\\
   +&\frac{0.000921 \alpha_0 ^4  W_0^4}{M^4}-\frac{0.185879 \alpha_0 ^2 \beta_0  W_0^4}{M^4}+\frac{0.000059 \alpha_0 ^2  W_0^4}{M^4}\nn\\
   -&\frac{0.029101 \alpha_0  \gamma_0  W_0^4}{M^4}-\frac{0.022607 \beta_0 ^2  W_0^4}{M^4}-\frac{0.001943 \beta_0  W_0^4}{M^4}\nn\\
  -&\frac{0.002777 \delta_0   W_0^4}{M^4}-\frac{0.004428 W_0^4}{M^4}+\frac{0.094989 \alpha_0 ^2 W_0  W_2}{M^4}\nn\\
   +&\frac{0.0326777 \beta_0  W_0 W_2}{M^4}+\frac{0.005692 W_0  W_2}{M^4} 
+\ldots\Big) 
\end{align}
which in terms of circumferential radius can be written as,

\begin{align} R_{\textrm{{\tiny ISCO}}}=&6M\Big(1+\frac{0.452733 \alpha_0  W_0}{M}-\frac{0.544331 J_1}{M^2}+\frac{0.040133 \alpha_0 ^2 W_0^2}{M^2}\nn\\
-&\frac{0.053788 \beta_0  W_0^2}{M^2}-\frac{0.059668 W_0^2}{M^2}-\frac{0.110743 \alpha_0  J_1 W_0}{M^3}\nn\\
   -&\frac{0.179922 M_2}{M^3}-\frac{0.011388 \alpha_0 ^3  W_0^3}{M^3}-\frac{0.159112 \alpha_0  \beta_0  W_0^3}{M^3}\nn\\
   +&\frac{0.017135 \alpha_0  W_0^3}{M^3}-\frac{0.018819 \gamma_0  W_0^3}{M^3}\nn\\
   +&\frac{0.054768 \alpha_0  W_2}{M^3}+\frac{0.054505 J_3}{M^4}-\frac{0.226244 J_1^2}{M^4}\nn\\
   -&\frac{0.138986 \alpha_0 ^2 J_1   W_0^2}{M^4}-\frac{0.011282 \beta_0  J_1 W_0^2}{M^4}\nn\\
   -&\frac{0.054650 J_1   W_0^2}{M^4}-\frac{0.002465 \alpha_0  M_2 W_0}{M^4}\nn\\
   +&\frac{0.001609 \alpha_0 ^4  W_0^4}{M^4}-\frac{0.14352 \alpha_0 ^2 \beta_0  W_0^4}{M^4}\nn\\
   +&\frac{0.000237 \alpha_0 ^2  W_0^4}{M^4}-\frac{0.023304 \alpha_0  \gamma_0  W_0^4}{M^4}\nn\\
   -&\frac{0.015496 \beta_0 ^2  W_0^4}{M^4}-\frac{0.000155 \beta_0  W_0^4}{M^4}\nn\\
   -&\frac{0.002151 \delta_0    W_0^4}{M^4}-\frac{0.003643 W_0^4}{M^4}\nn\\
   +&\frac{0.075878 \alpha_0 ^2 W_0  W_2}{M^4}+\frac{0.025303 \beta_0  W_0 W_2}{M^4}\nn\\
   +&\frac{0.004590 W_0   W_2}{M^4}
+\dots\Big)
   \end{align}
Similarly, the orbital frequency at the ISCO is given, up to fourth order, by the expression 
\begin{align}
\Omega_{\textrm{{\tiny ISCO}}}=&\frac{1}{6 \sqrt{6} M}\Big( 1 -\frac{0.75 \alpha_0  W_0}{M}+\frac{0.748455 J_1}{M^2}
-\frac{0.000218 \alpha_0 ^2 W_0^2}{M^2}\nn\\
+&\frac{0.035481 \beta_0    W_0^2}{M^2}+\frac{0.077687 W_0^2}{M^2} +\frac{0.23433 M_2}{M^3}\nn\\
   -&\frac{0.509757 \alpha_0  J_1 W_0}{M^3}-\frac{0.00784 \alpha_0 ^3   W_0^3}{M^3}+\frac{0.117236 \alpha_0  \beta_0  W_0^3}{M^3}\nn\\
   -&\frac{0.089082 \alpha_0    W_0^3}{M^3}+\frac{0.022604 \gamma_0  W_0^3}{M^3}-\frac{0.065506 \alpha_0   W_2}{M^3}\nn\\
   +&\frac{0.780664 J_1^2}{M^4}-\frac{0.204041 \alpha_0  M_2 W_0}{M^4}-\frac{0.0425199 \alpha_0 ^4  W_0^4}{M^4}\nn\\
   +&\frac{0.009988 \alpha_0 ^2 \beta_0  W_0^4}{M^4}+\frac{0.022952 \alpha_0 ^2  W_0^4}{M^4}\nn\\
   +&\frac{0.009234 \alpha_0  \gamma_0  W_0^4}{M^4}+\frac{0.017918 \beta_0 ^2  W_0^4}{M^4}+\frac{0.006122 \beta_0  W_0^4}{M^4}\nn\\
   +&\frac{0.002812 \delta_0   W_0^4}{M^4}+\frac{0.009612 W_0^4}{M^4}-\frac{0.037394 \alpha_0 ^2 W_0 W_2}{M^4}\nn\\
   -&\frac{0.033029 \beta_0  W_0 W_4}{M^4}-\frac{0.006338 W_0  W_2}{M^4}\nn\\
   +&\frac{0.114417 \alpha_0 ^2 J_1  W_0^2}{M^4}+\frac{0.075325 \beta_0  J_1 W_0^2}{M^4}\nn\\
   +&\frac{0.168052 J_1  W_0^2}{M^4}
+ \ldots \Big).
 \end{align}

If we now use the second scaling for the multipole moments, the resulting expression up to fourth order for the ISCO is,

 \begin{align} \rho_{\textrm{{\tiny ISCO}}}=& 2 \sqrt{6} M \Big(1+\frac{5 \alpha_0  W_0}{16 M}  -\frac{5 J_1}{3  \sqrt{6} M^2} +\frac{0.037298 \alpha_0 ^2 W_0^2}{M^2}\nn\\
   -&\frac{0.092923 \beta_0   W_0^2}{M^2}-\frac{0.072877 W_0^2}{M^2}-\frac{0.21977 M_2}{M^3} \nn\\
   -&\frac{0.009463 \alpha_0  ^3 W_0^3}{M^3}-\frac{0.1979119 \alpha_0  \beta_0  W_0^3}{M^3}\nn\\
   +&\frac{0.019801 \alpha_0  W_0^3}{M^3}-\frac{0.02526 \gamma_0  W_0^3}{M^3}\nn\\
   +&\frac{0.073633 \alpha_0  W_2}{M^3}-\frac{0.162089 \alpha_0  J_1 W_0}{M^3}\nn\\
   +&\frac{0.068131 J_3}{M^4}-\frac{0.289247 J_1^2}{M^4}-\frac{0.013148 \alpha_0  M_2 W_0}{M^4}\nn\\
   +&\frac{0.000921 \alpha_0 ^4 W_0^4}{M^4}-\frac{0.185879 \alpha_0 ^2 \beta_0  W_0^4}{M^4}+\frac{0.000059 \alpha_0  ^2 W_0^4}{M^4}\nn\\
   -&\frac{0.029101 \alpha_0  \gamma_0  W_0^4}{M^4}-\frac{0.022607 \beta_0 ^2 W_0^4}{M^4}\nn\\
   -&\frac{0.001943 \beta_0  W_0^4}{M^4}-\frac{0.002777 \delta_0  W_0^4}{M^4}\nn\\
   -&\frac{0.004428 W_0^4}{M^4}+\frac{0.094989 \alpha_0 ^2 W_0 W_2}{M^4}\nn\\
   +&\frac{0.032678 \beta_0  W_0 W_2}{M^4}+\frac{0.005692 W_0 W_2}{M^4}\nn\\
   -&\frac{0.181816 \alpha_0 ^2 J_1 W_0^2}{M^4}-\frac{0.037135  \beta_0  J_1 W_0^2}{M^4}\nn\\
   -&\frac{0.067825 J_1  W_0^2}{M^4}
 +\ldots \Big)
%
 \end{align}
which in terms of circumferential radius can be written as,
 \begin{align}  R_{\textrm{{\tiny ISCO}}}=& 6 M \Big(1+\frac{0.452733 \alpha_0  W_0}{M}  -\frac{0.544331 J_1}{M^2}\nn\\
  +&\frac{0.040133 \alpha_0 ^2 W_0^2}{M^2}-\frac{0.053788 \beta_0  W_0^2}{M^2}-\frac{0.059668  W_0^2}{M^2}\nn\\
   -&\frac{0.110743 \alpha_0  J_1 W_0}{M^3}  -\frac{0.179922 M_2}{M^3}-\frac{0.011388 \alpha_0 ^3 W_0^3}{M^3}\nn\\
   -&\frac{0.159112 \alpha_0  \beta_0    W_0^3}{M^3}+\frac{0.017135 \alpha_0  W_0^3}{M^3}\nn\\
   -&\frac{0.018819 \gamma_0   W_0^3}{M^3}+\frac{0.054768 \alpha_0  W_2}{M^3}\nn\\
   +&\frac{0.054505  J_3}{M^4}-\frac{0.138986 \alpha_0 ^2 J_1 W_0^2}{M^4}-\frac{0.011282 \beta_0  J_1 W_0^2}{M^4}\nn\\
   -&\frac{0.05465 J_1  W_0^2}{M^4} -\frac{0.226244 J_1^2}{M^4}\nn\\ 
   -&\frac{0.002465 \alpha_0  M_2 W_0}{M^4}+\frac{0.001609 \alpha_0 ^4 W_0^4}{M^4}\nn\\
   -&\frac{0.14352 \alpha_0 ^2 \beta_0  W_0^4}{M^4}+\frac{0.000237 \alpha_0 ^2 W_0^4}{M^4}\nn\\
   -&\frac{0.023304 \alpha_0  \gamma_0 W_0^4}{M^4}-\frac{0.015496 \beta_0 ^2 W_0^4}{M^4}\nn\\
   -&\frac{0.000155 \beta_0 W_0^4}{M^4}-\frac{0.002151 \delta_0  W_0^4}{M^4}\nn\\
   -&\frac{0.003643 W_0^4}{M^4}+\frac{0.075878 \alpha_0 ^2 W_0 W_2}{M^4}\nn\\
   +&\frac{0.025303 \beta_0  W_0 W_2}{M^4}+\frac{0.00459 W_0 W_2}{M^4}
 +\ldots \Big)
 \end{align} 
Finally, the orbital frequency at the ISCO is given, up to fourth order, by the expression 
  \begin{align} 
\Omega_{\textrm{{\tiny ISCO}}}=&\frac{1}{6 \sqrt{6} M}\Big( 1 -\frac{0.75 \alpha_0  W_0}{M}+\frac{0.748455 J_1}{M^2}\nn\\
  -&\frac{0.000218 \alpha_0 ^2 W_0^2}{M^2}+\frac{0.035481 \beta_0  W_0^2}{M^2}\nn\\
   +&\frac{0.077688 W_0^2}{M^2} +\frac{0.23433 M_2}{M^3}-\frac{0.00784 \alpha_0 ^3 W_0^3}{M^3}\nn\\
   +&\frac{0.117236 \alpha_0  \beta_0  W_0^3}{M^3}-\frac{0.089082 \alpha_0  W_0^3}{M^3}\nn\\
   +&\frac{0.022604 \gamma_0  W_0^3}{M^3}-\frac{0.065506 \alpha_0   W_2}{M^3}\nn\\
   -&\frac{0.509757 \alpha_0  J_1 W_0}{M^3} +\frac{0.780664 J_1^2}{M^4}-\frac{0.076216   J_3}{M^4}\nn\\
   -&\frac{0.204041 \alpha_0  M_2 W_0}{M^4}-\frac{0.04252 \alpha_0 ^4 W_0^4}{M^4}\nn\\
   +&\frac{0.009988 \alpha_0 ^2 \beta_0  W_0^4}{M^4}+\frac{0.022952 \alpha_0 ^2 W_0^4}{M^4}\nn\\
   +&\frac{0.009234 \alpha_0  \gamma_0  W_0^4}{M^4}+\frac{0.017918 \beta_0 ^2 W_0^4}{M^4}\nn\\
   +&\frac{0.006122\beta_0  W_0^4}{M^4}+\frac{0.002812 \delta_0   W_0^4}{M^4}\nn\\
   +&\frac{0.009612 W_0^4}{M^4}-\frac{0.037394 \alpha_0 ^2 W_0 W_2}{M^4}\nn\\
   -&\frac{0.033029 \beta_0  W_0 W_2}{M^4}-\frac{0.006338 W_0 W_2}{M^4}\nn\\
   +&\frac{0.114417 \alpha_0 ^2 J_1 W_0^2}{M^4}+\frac{0.075325 \beta_0  J_1 W_0^2}{M^4}\nn\\
   +&\frac{0.168052 J_1 W_0^2}{M^4}
+\ldots \Big).
 \end{align} 

It is straightforward to verify that these expression reduce to the expressions of the spherical case, once the higher order moments of the spacetime and of the scalar field are set to zero (modulo variations due to numerical accuracy).

\section{Comparing GR to scalar-tensor gravity}
\label{sec:application}

The comparison of the expressions that we have produced here with the corresponding expressions in GR, lead to some interesting conclusions regarding the phenomenology of a compact star in scalar-tensor theory.

The first thing that we should examine is the leading order behaviour of the orbital frequency of circular orbits and how this compares to the behaviour in GR. One can see that to leading order, the orbital frequency behaves as 
\be \Omega \sim \left(\frac{M-W_0 \alpha_0}{\rho^3}\right)^{1/2}.\ee
This is the  frequency one would measure for an object orbiting far from the central compact object, {\em i.e.}~the ``Keplerian'' orbital frequency. So, if we were to measure the orbital motion of matter orbiting far from the compact object in order to estimate its mass, the mass that we would infer assuming ``Keplerian'' motion would be $\bar{M}=M-W_0 \alpha_0$. It is the quantity $\bar{M}$ therefore that is the observable mass of the scalar-tensor theory and it is the mass of the compact object in the same sense as we would infer the mass in GR.

Starting from that, we can attempt to compare the expressions for $\Delta\tilde{E}$, $\Omega_{\rho}/\Omega$ and $\Omega_z/\Omega$ order by order in $U$ and see if we can distinguish the orbital behaviour of particles between scalar-tensor gravity and GR. The first three terms of $\Delta\tilde{E}$ in GR read,
\be \Delta \tilde{E}=
\frac{1}{3}U^2-\frac{1}{2}U^4 +\frac{20 J_1 }{9 M^2}U^5+O\left(U^6\right),
\ee
where $U=(M\Omega)^{1/3}$, with $M$ being the mass as infered from Keplerian orbits in GR. The corresponding terms in scalar-tensor theory, as we have seen in Section 4, are 
\bea \Delta \tilde{E}\!\!\!\!\!\!&=&\!\!\!\!\!\!
\frac{1}{3}U^2+\left(\frac{2 \beta_0  W_0^2}{9 \bar{M}^2}-\frac{8 \alpha_0  W_0}{9
   \bar{M}}-\frac{1}{2}\right)U^4 +\frac{20 J_1 }{9 \bar{M}^2}U^5\nn\\
   \!\!\!\!\!\!&&\!\!\!\!\!\!+O\left(U^6\right),
\eea
where $U=(\bar{M}\Omega)^{1/3}$ for the reasons discussed above.
The comparison between the two expressions shows that, even when we can't tell observationally if the mass that we are measuring corresponds to the GR $M$ or the scalar-tensor $\bar{M}$, we can still distinguish between GR and scalar-tensor theory. If we adopt the general form for the expansion of $\Delta\tilde{E}$ with respect to $U$ to be 
\be \Delta\tilde{E}=\sum_{k=2}^{\infty}A_k U^k, \ee
then the ratio 
\be \frac{A_4}{A_2}=\left \{
\begin{matrix}
-3/2&, & \textrm{GR} \\
\left(\frac{2 \beta_0  W_0^2}{3 \bar{M}^2}-\frac{8 \alpha_0  W_0}{3 \bar{M}}-\frac{3}{2}\right)&, & \textrm{scalar-tensor}
\end{matrix}\right.
\ee
could be used as a tool to constrain deviations from GR or measure the scalar charge of a compact object. Current constrains from solar system experiments and pulsar timing require that  $\alpha_0\leq10^{-3}$. However, even for $\alpha_0=0$ there is a discrepancy between GR and scalar-tensor theory that is equal to $2 \beta_0  W_0^2/3 M^2$. 
Interestingly enough this deviation from GR manifests itself at an order where the rotation has no contribution yet. Therefore it is present for both rotating and non-rotating stars and depends only on the scalar monopole and the parameter $\beta_0$ of the theory. To give a first estimate of the magnitude of the effect, we present in Table \ref{DeltaEtab} the relative difference of the ratio $A_4/A_2$ between GR and scalar-tensor theory, $\Delta(A_4/A_2)$, defined as the difference of the GR value minus the scalar-tensor value divided by the GR value. The scalarized neutron star models used for the table are the non-rotating models in the work by \cite{DonevaSTprecess2014}. They are generated using the APR equation of state (EOS) for $\beta_0=-4.5$ and $\phi_{\infty}=0$. 
We can see that  $\Delta(A_4/A_2)$ is in the range of $1\%-10\%$, and we expect it to be even larger for rotating neutron stars. 

\begin{table}  
\caption{Values for the mass $M$, the scalar charge $W_0$  and the resulting deviations from GR, for non-rotating neutron star models constructed with the APR EOS. The values were taken from  \protect\cite{DonevaSTprecess2014} 
and the models assumed $\beta_0=-4.5$, $\alpha_0=0$ and a value for the scalar field at infinity, $\phi_{\infty}=0$.}\protect \label{DeltaEtab}
 {\centering
 \begin{tabular}{c c c c c}\hline
  $M/M_{\odot}$ & $M$ & $W_0$ & $\Delta(A_4/A_2)$ & $\Delta B_2$  \\
   	&  (km)  &  (km)  & (per cent)  & (per cent) \\ \hline
 1.68 & 2.52 & -0.3596 & -4.07257 & -0.509071 \\
 1.77 & 2.655 & -0.4868 & -6.7236 & -0.84045 \\
 1.86 & 2.79 & -0.5448 & -7.62598 & -0.953248 \\
 1.94 & 2.91 & -0.5415 & -6.92534 & -0.865667 \\
 2 & 3. & -0.4608 & -4.71859 & -0.589824 \\
 2.05 & 3.075 & -0.2414 & -1.23258 & -0.154072 \\ \hline
 \end{tabular}
}
\end{table}

The same analysis can be performed for the expansion of $\Omega_{\rho}/\Omega$. The first terms of the expansion in GR read 
\be \frac{\Omega_{\rho}}{\Omega}=3U^2 -\frac{4 J_1 }{M^2} U^3+O\left(U^4\right),
   \ee
while the corresponding terms in scalar-tensor theory are,
\be \frac{\Omega_{\rho}}{\Omega}=
\left(3-\frac{W_0 \left(\beta_0  W_0-8 \alpha_0  \bar{M}\right)}{2 \bar{M}^2}\right)U^2 -\frac{4
   J_1 }{\bar{M}^2} U^3  +O\left(U^4\right).
   \ee
One can see that the periastron precession at leading order behaves as $\Omega_{\rho}/\Omega\sim 3U^2$ in GR while in scalar-tensor this is modified. If we assume the expansion for the periastron precession, 
\be \Omega_{\rho}/\Omega=\sum_{k=2}^{\infty}B_k U^k, \ee
then the relative difference between GR and scalar-tensor theory 
at leading order will be,
\be \Delta B_2 =\frac{W_0 \left(\beta_0  W_0-8 \alpha_0  \bar{M}\right)}{6 \bar{M}^2}. \ee
Again if we impose the observational constrains that reduce $\alpha_0$ to be almost zero, we can see that the discrepancy between GR and scalar-tensor theory remains (although it is smaller than in the case of the ratio $(A_4/A_2)$ by a factor of 8). Using the same numerical examples as before, we estimate in Table \ref{DeltaEtab} the magnitude of the relative difference in $B_2$ between GR and scalar-tensor theory.

The case of the nodal precession is slightly different, because of the strict rotation dependence of the effect. The differences between GR and scalar-tensor theory comes about at higher order with respect to the multipole moments. In GR the nodal precession expansion has the form, 
\be \frac{\Omega_z}{\Omega}=
\frac{2 J_1 }{M^2}U^3+\frac{3  M_2}{2 M^3} U^4+O\left(U^6\right),
\ee
where we should note that the next term after the term of order $U^4$ in the expansion, is of order $U^6$. In contrast, the nodal precession frequency in scalar-tensor theory has the expansion
\begin{align}
 \frac{\Omega_z}{\Omega} =&
\frac{2 J_1 }{\bar{M}^2}U^3 \!+\! \frac{3 (M_2 \!-\! \alpha_0  W_2)}{2 \bar{M}^3}U^4  \nn\\
&-  \frac{2 J_1
    W_0 \left(\beta_0  W_0-\alpha_0  \bar{M}\right)}{\bar{M}^4}U^5 +O\left(U^6\right),
\end{align}
where apart from the differences with respect to the $U^4$ term, there is also a term of order $U^5$. If we impose the observetional constrains for $\alpha_0$ in this case, we can see that the terms of order $U^3$ and $U^4$ are the same between GR and scalar-tensor theory, but the $U^5$ term  is non-zero in scalar-tensor. Unfortunately, this discrepancy probably comes at a high enough order to make measuring it in observations quite challenging. 

\section{Conclusions and Outlook}
\label{sec:conclusions}

We have used the recently developed multipole moment formalism in scalar tensor theory in order to obtain expressions for various observables that characterise geodesics in terms of the moments. 
Ryan had derived similar expressions in GR with the intention to use them for extreme mass ratio inspirals (EMRIs) as a tool to measure the multipole moments of the spacetime around a supermassive BHs from gravitational waves observations. It is questionable whether the expressions we derived here will be useful in this particular scenario in scalar tensor theory. This is because no-hair theorems  by \cite{Hawking:1972qk,Bekenstein1972PhRvD5,Bekenstein1995PhRvD51,Mayo1996PhRvD54,Sotiriou:2011dz}, suggest that BHs will not carry a scalar charge. Realistic astrophysical BHs might circumvent no-hair theorems by violating one or more of their assumptions, see a recent review by \cite{Sotiriou:2015pka} for a discussion.  For instance, the presence of matter around the BH can introduce a scalar charge (black hole scalarization) as it was shown by \cite{Cardoso:2013fwa,Cardoso:2013opa}. Whether such a charge would be large enough to lead to some measurable effect is not clear at this stage. 

In principle one could consider using the relations between the observables and the multipole moments in EMRIs as a direct test of the BH scalarization hypothesis. Something similar could be done with supermassive BHs or stelar mass BHs using QPOs. After the first discovery by \cite{AGNQPO2008Natur} of QPOs in an active galaxy, several more sources have been verified [see for example the review by \cite{AGNQPOsRev2014}]. These QPO sources could be used to test whether supermassive BHs are scalarized by implementing the expressions for the frequencies presented here. Additionally to these sources, \cite{JohanPsaltis2011ApJ} have argued that Sgr A* could be another promising source of QPOs, measured using very-long baseline interferometry, that could be used for this sort of tests.   
 Apart from supermassive BHs, QPOs are commonly observed in X-ray binaries that host stellar mass BHs (see \cite{lamb,derKlis}). 
Therefore, with improved future observations it might be possible to test the BH scalarization scenario.  

Since it is still questionable whether BHs can actually carry any significant scalar charge in scalar-tensor theory, the primary class of systems in which one could apply the derived relations are those when the central object is a NS. Therefore NSs in LMXBs are probably the most promising sources for implementing the expressions developed here. These  are usually members of a binary system, which provides the opportunity for an independent relatively accurate measurement of the Keplerian mass. This could make even more successful the implementation of the extended Ryan expressions, since one would have less fitting parameters (the mass would be independently known) if one were to attempt a fit of the QPOs frequencies along the lines that was proposed by \cite{PappasQPOs} (additionally one would need to fit lower order coefficients than those discussed in the latter work).

Furthermore one could attempt to associate the accretion spectrum and the disc temperature distribution, to the expression for $\Delta\tilde{E}$ in hope of probing the deviation between scalar-tensor theory and GR using some sort of disc tomography or the iron line reverberation technic (see for example work by \cite{Jiang:2014loa}).    
 
 It should be noted that since we considered here only observables associated with geodesics of the metric, our approach does not actually account for the effect of a scalar-scalar interaction between the central objects and the orbiting matter. We have essentially worked under the assumption that the orbiting matter is not scalarized, otherwise it would not follow geodesics of the metric in the first place. 
 
Further analysis is needed to explore the possibilities for using the expressions we have derived in order to extract constrains for scalar-tensor theories as a deviation from GR. This would require detailed modelling of the different classes of sources that we have discussed here and an artful consideration of current and future observational uncertainties.

\section*{Acknowledgements}

The authors would like to thank P. Pani and H. O. da Silva for sharing with us data regarding scalarized stars. The research leading to these results has received funding from the European Research Council under the European Union's Seventh Framework Programme (FP7/2007-2013) / ERC Grant Agreement n.~306425 ``Challenging General Relativity''. Research at Perimeter Institute is supported by the Government of Canada through Industry Canada and by the Province of Ontario through the Ministry of Economic Development \& Innovation.

\bibliographystyle{mn} 
\bibliography{mn-jour,mybibliography}

\newpage
\appendix

%
\section[]{Full expressions for the ISCO}
\label{sec:fullISCO}

Here we present the complete expansion for the ISCO radii and the frequencies discussed in Sec.~\ref{sec:isco}.%

\subsection[]{Scalar moments scaling as, $W_2\sim\varepsilon^3$, and $W_4\sim\varepsilon^5$}

\begin{align} \rho_{\textrm{{\tiny ISCO}}}=& 2 \sqrt{6} M \Big(1+\frac{5
   \alpha_0  W_0}{16 M}-\frac{5 J_1}{3 \sqrt{6} M^2}+\frac{0.0372978 \alpha_0 ^2
   W_0^2}{M^2}\nn\\
   &-\frac{0.092923 \beta_0  W_0^2}{M^2}-\frac{0.072877 W_0^2}{M^2}-\frac{0.21977 M_2}{M^3}\nn\\
   &-\frac{0.009463 \alpha_0 ^3 W_0^3}{M^3}-\frac{0.197912 \alpha_0  \beta_0 
   W_0^3}{M^3}\nn\\
   &+\frac{0.019801 \alpha_0  W_0^3}{M^3}-\frac{0.025259 \gamma_0 
   W_0^3}{M^3}+\frac{0.073633 \alpha_0  W_2}{M^3}\nn\\
   &-\frac{0.162089 \alpha_0 
   J_1 W_0}{M^3}+\frac{0.068131 J_3}{M^4}-\frac{0.289247
   J_1^2}{M^4}\nn\\
   &-\frac{0.013148 \alpha_0  M_2
   W_0}{M^4}-\frac{0.181816 \alpha_0 ^2 J_1 W_0^2}{M^4}\nn\\
   &-\frac{0.037135 \beta_0 
   J_1 W_0^2}{M^4}-\frac{0.0678245 J_1 W_0^2}{M^4}\nn\\
   &+\frac{0.000921 \alpha_0 ^4
   W_0^4}{M^4}-\frac{0.185879 \alpha_0 ^2 \beta_0  W_0^4}{M^4}+\frac{0.000059 \alpha_0 ^2
   W_0^4}{M^4}\nn\\
   &-\frac{0.029101 \alpha_0  \gamma_0  W_0^4}{M^4}-\frac{0.022607 \beta_0 ^2
   W_0^4}{M^4}-\frac{0.001943 \beta_0  W_0^4}{M^4}\nn\\
  & -\frac{0.002777 \delta_0 
   W_0^4}{M^4}-\frac{0.004428 W_0^4}{M^4}+\frac{0.094989 \alpha_0 ^2 W_0
   W_2}{M^4}\nn\\
   &+\frac{0.0326777 \beta_0  W_0 W_2}{M^4}+\frac{0.005692 W_0
   W_2}{M^4}+\frac{0.019922 M_4}{M^5}\nn\\
   &-\frac{0.329286 J_1
   M_2}{M^5}-\frac{0.010686 \alpha_0  J_1^2 W_0}{M^5}-\frac{0.0244395 \alpha_0 
   J_3 W_0}{M^5}\nn\\
   &-\frac{0.095249 \alpha_0 ^2 M_2 W_0^2}{M^5}-\frac{0.038733 \beta_0  M_2
   W_0^2}{M^5}-\frac{0.033822 M_2 W_0^2}{M^5}\nn\\
   &-\frac{0.171355 \alpha_0 ^3 J_1W_0
   W_0^3}{M^5}-\frac{0.140523 \alpha_0  \beta_0  J_1 W_0^3}{M^5}\nn\\
   &+\frac{0.024257 \alpha_0  J_1
   W_0^3}{M^5}-\frac{0.02547 \gamma_0  J_1 W_0^3}{M^5}\nn\\
   &+\frac{0.070744 \alpha_0  J_1
   W_2}{M^5} -\frac{0.076987
   M_2^2}{M^6}+\frac{0.149635 J_1 J_3}{M^6}\nn\\
   &-\frac{0.296483 
   J_1^3}{M^6} +\frac{0.049323 \alpha_0  J_1 M_2 W_0}{M^6} -\frac{0.124683 \alpha_0 ^2 J_1^2
   W_0^2}{M^6}\nn\\
   &-\frac{0.057378 \beta_0  J_1^2 W_0^2}{M^6}-\frac{0.097335 J_1^2
   W_0^2}{M^6}+\frac{0.037166 \alpha_0 J_1^3 W_0}{M^7}\nn\\
   &-\frac{0.568603 J_1^2 M_2}{M^7}-\frac{0.384317 J_1^4}{M^8}+\ldots\Big) 
\end{align}

\begin{align} R_{\textrm{{\tiny ISCO}}}=&6M\Big(1+\frac{0.452733 \alpha_0  W_0}{M}-\frac{0.544331 J_1}{M^2}+\frac{0.040133 \alpha_0 ^2 W_0^2}{M^2}\nn\\
&-\frac{0.053788 \beta_0 
   W_0^2}{M^2}-\frac{0.059668 W_0^2}{M^2}-\frac{0.110743 \alpha_0  J_1 W_0}{M^3}\nn\\
   &-\frac{0.179922 M_2}{M^3}-\frac{0.011388 \alpha_0 ^3
   W_0^3}{M^3}-\frac{0.159112 \alpha_0  \beta_0  W_0^3}{M^3}\nn\\
   &+\frac{0.017135 \alpha_0 
   W_0^3}{M^3}-\frac{0.018819 \gamma_0  W_0^3}{M^3}\nn\\
   &+\frac{0.054768 \alpha_0 
   W_2}{M^3}+\frac{0.054505 J_3}{M^4}-\frac{0.226244 J_1^2}{M^4}\nn\\
   &-\frac{0.138986 \alpha_0 ^2 J_1
   W_0^2}{M^4}-\frac{0.011282 \beta_0  J_1 W_0^2}{M^4}\nn\\
   &-\frac{0.054650 J_1
   W_0^2}{M^4}-\frac{0.002465 \alpha_0  M_2 W_0}{M^4}\nn\\
   &+\frac{0.001609 \alpha_0 ^4
   W_0^4}{M^4}-\frac{0.14352 \alpha_0 ^2 \beta_0  W_0^4}{M^4}\nn\\
   &+\frac{0.000237 \alpha_0 ^2
   W_0^4}{M^4}-\frac{0.023304 \alpha_0  \gamma_0  W_0^4}{M^4}\nn\\
   &-\frac{0.015496 \beta_0 ^2
   W_0^4}{M^4}-\frac{0.000155 \beta_0  W_0^4}{M^4}\nn\\
   &-\frac{0.002151 \delta_0 
   W_0^4}{M^4}-\frac{0.003643 W_0^4}{M^4}\nn\\
   &+\frac{0.075878 \alpha_0 ^2 W_0
   W_2}{M^4}+\frac{0.025303 \beta_0  W_0 W_2}{M^4}\nn\\
   &+\frac{0.004590 W_0
   W_2}{M^4}-\frac{0.0214468
   \alpha_0  J_3 W_0}{M^5}\nn\\
   &-\frac{0.264597 J_1
   M_2}{M^5}+\frac{0.0160475 \alpha_0  J_1^2 W_0}{M^5}\nn\\
   &-\frac{0.135488 \alpha_0 ^3 J_1 W_0^3}{M^5}-\frac{0.105171 \alpha_0  \beta_0  J_1
   W_0^3}{M^5}\nn\\
   &+\frac{0.022624 \alpha_0  J_1 W_0^3}{M^5}-\frac{0.017785 \gamma_0  J_1
   W_0^3}{M^5}\nn\\
   &+\frac{0.048952 \alpha_0  J_1 W_2}{M^5}-\frac{0.075539 \alpha_0 ^2 M_2 W_0^2}{M^5}\nn\\
   &-\frac{0.025925 \beta_0  M_2
   W_0^2}{M^5}-\frac{0.027709 M_2 W_0^2}{M^5}\nn\\
   &+\frac{0.016066
   M_4}{M^5}-\frac{0.062736
   M_2^2}{M^6}-\frac{0.230289
   J_1^3}{M^6}\nn\\
   &+\frac{0.118328
   J_1 J_3}{M^6}+\frac{0.0602789 \alpha_0  J_1 M_2 W_0}{M^6}\nn\\
   &-\frac{0.090747 \alpha_0 ^2 J_1^2
   W_0^2}{M^6}-\frac{0.0231068 \beta_0  J_1^2 W_0^2}{M^6}\nn\\
   &-\frac{0.077521 J_1^2
   W_0^2}{M^6}-\frac{0.451308 J_1^2 M_2}{M^7}+\frac{0.062687 \alpha_0  J_1^3 W_0}{M^7}\nn\\
   &-\frac{0.297466 J_1^4}{M^8}+\dots\Big)
   \end{align}

\begin{align}
\Omega_{\textrm{{\tiny ISCO}}}=&\frac{1}{6 \sqrt{6} M}\Big( 1 -\frac{0.75 \alpha_0  W_0}{M}+\frac{0.748455 J_1}{M^2}
-\frac{0.000218 \alpha_0 ^2 W_0^2}{M^2}\nn\\
&+\frac{0.035481 \beta_0 
   W_0^2}{M^2}+\frac{0.077687 W_0^2}{M^2} +\frac{0.23433 M_2}{M^3}\nn\\
   &-\frac{0.509757 \alpha_0  J_1 W_0}{M^3}-\frac{0.00784 \alpha_0 ^3
   W_0^3}{M^3}+\frac{0.117236 \alpha_0  \beta_0  W_0^3}{M^3}\nn\\
   &-\frac{0.089082 \alpha_0 
   W_0^3}{M^3}+\frac{0.022604 \gamma_0  W_0^3}{M^3}-\frac{0.065506 \alpha_0 
   W_2}{M^3}\nn\\
   &+\frac{0.780664 J_1^2}{M^4}-\frac{0.204041 \alpha_0  M_2 W_0}{M^4}-\frac{0.0425199 \alpha_0 ^4
   W_0^4}{M^4}\nn\\
   &+\frac{0.009988 \alpha_0 ^2 \beta_0  W_0^4}{M^4}+\frac{0.022952 \alpha_0 ^2
   W_0^4}{M^4}\nn\\
   &+\frac{0.009234 \alpha_0  \gamma_0  W_0^4}{M^4}+\frac{0.017918 \beta_0 ^2
   W_0^4}{M^4}+\frac{0.006122 \beta_0  W_0^4}{M^4}\nn\\
   &+\frac{0.002812 \delta_0 
   W_0^4}{M^4}+\frac{0.009612 W_0^4}{M^4}-\frac{0.037394 \alpha_0 ^2 W_0
   W_2}{M^4}\nn\\
   &-\frac{0.033029 \beta_0  W_0 W_4}{M^4}-\frac{0.006338 W_0
   W_2}{M^4}\nn\\
   &+\frac{0.114417 \alpha_0 ^2 J_1
   W_0^2}{M^4}+\frac{0.075325 \beta_0  J_1 W_0^2}{M^4}\nn\\
   &+\frac{0.168052 J_1
   W_0^2}{M^4}+\frac{0.646435 J_1
   M_2}{M^5}-\frac{0.55815 \alpha_0  J_1^2 W_0}{M^5}\nn\\
   &+\frac{0.04601 \alpha_0 ^3 J_1 W_0^3}{M^5}+\frac{0.260446 \alpha_0  \beta_0  J_1
   W_0^3}{M^5}\nn\\
   &-\frac{0.190736 \alpha_0  J_1 W_0^3}{M^5}+\frac{0.05127 \gamma_0  J_1
   W_0^3}{M^5}\nn\\
   &-\frac{0.144884 \alpha_0  J_1 W_2}{M^5}+\frac{0.093943
   \alpha_0  J_3 W_0}{M^5}-\frac{0.076216 J_3}{M^4}\nn\\
   &+\frac{0.107837 \alpha_0 ^2 M_2 W_0^2}{M^5}+\frac{0.051719 \beta_0  M_2
   W_0^2}{M^5}\nn\\
   &+\frac{0.065906 M_2 W_0^2}{M^5}-\frac{0.022183
   M_4}{M^5}+\frac{0.12833
   M_2^2}{M^6}\nn\\
   &+\frac{0.981402
   J_1^3}{M^6}-\frac{0.260058 J_1
   J_3}{M^6}\nn\\
   &-\frac{0.605842 \alpha_0  J_1 M_2 W_0}{M^6}+\frac{0.287271 \alpha_0 ^2 J_1^2
   W_0^2}{M^6}\nn\\
   &+\frac{0.147587 \beta_0  J_1^2 W_0^2}{M^6}+\frac{0.315357 J_1^2
   W_0^2}{M^6}\nn\\
   &-\frac{0.782788 \alpha_0  J_1^3 W_0}{M^7}+\frac{1.42856 J_1^2 M_2}{M^7}
   \nn\\
   &+\frac{1.38245 J_1^4}{M^8}
+ \ldots \Big).
 \end{align}

\subsection[]{Scalar moments scaling as, $W_2\sim\varepsilon^2$, and $W_4\sim\varepsilon^4$}

 \begin{align} \rho_{\textrm{{\tiny ISCO}}}=&
 2 \sqrt{6} M \Big(1+\frac{5 \alpha_0  W_0}{16
   M}  -\frac{5 J_1}{3
   \sqrt{6} M^2} 
   +\frac{0.037298 \alpha_0 ^2 W_0^2}{M^2}\nn\\
   &-\frac{0.092923 \beta_0 
   W_0^2}{M^2}-\frac{0.072877 W_0^2}{M^2}-\frac{0.21977 M_2}{M^3} \nn\\
 &  -\frac{0.009463 \alpha_0
   ^3 W_0^3}{M^3}-\frac{0.1979119 \alpha_0  \beta_0  W_0^3}{M^3}\nn\\
   &+\frac{0.019801
   \alpha_0  W_0^3}{M^3}-\frac{0.02526 \gamma_0  W_0^3}{M^3}\nn\\
   &+\frac{0.073633
   \alpha_0  W_2}{M^3}-\frac{0.162089 \alpha_0  J_1 W_0}{M^3}\nn\\
 &   +\frac{0.068131 J_3}{M^4}-\frac{0.289247 J_1^2}{M^4}-\frac{0.013148
   \alpha_0  M_2 W_0}{M^4}\nn\\
   &+\frac{0.000921 \alpha_0 ^4
   W_0^4}{M^4}-\frac{0.185879 \alpha_0 ^2 \beta_0  W_0^4}{M^4}+\frac{0.000059 \alpha_0
   ^2 W_0^4}{M^4}\nn\\
   &-\frac{0.029101 \alpha_0  \gamma_0  W_0^4}{M^4}-\frac{0.022607 \beta_0
   ^2 W_0^4}{M^4}\nn\\
   &-\frac{0.001943 \beta_0  W_0^4}{M^4}-\frac{0.002777 \delta_0 
   W_0^4}{M^4}\nn\\
   &-\frac{0.004428 W_0^4}{M^4}+\frac{0.094989 \alpha_0 ^2
   W_0 W_2}{M^4}\nn\\
   &+\frac{0.032678 \beta_0  W_0
   W_2}{M^4}+\frac{0.005692 W_0 W_2}{M^4}\nn\\
   &-\frac{0.181816 \alpha_0 ^2 J_1 W_0^2}{M^4}-\frac{0.037135
   \beta_0  J_1 W_0^2}{M^4}\nn\\
   &-\frac{0.067825 J_1
   W_0^2}{M^4}+\frac{0.019922
   M_4}{M^5}-\frac{0.329286 J_1
   M_2}{M^5}\nn\\
   &-\frac{0.095249 \alpha_0 ^2 M_2 W_0^2}{M^5}-\frac{0.038734 \beta_0 
   M_2 W_0^2}{M^5}\nn\\
   &-\frac{0.033822 M_2 W_0^2}{M^5} -\frac{0.171355 \alpha_0 ^3
   J_1 W_0^3}{M^5}-\frac{0.140523 \alpha_0  \beta_0  J_1
   W_0^3}{M^5}\nn\\
   &+\frac{0.024257 \alpha_0  J_1 W_0^3}{M^5}-\frac{0.02547 \gamma_0 
   J_1 W_0^3}{M^5}\nn\\
   &+\frac{0.070744 \alpha_0  J_1
   W_2}{M^5}-\frac{0.010686 \alpha_0 
   J_1^2 W_0}{M^5}\nn\\
   &   -\frac{0.0244395 \alpha_0  J_3 W_0}{M^5}+\frac{0.077723 \alpha_0 ^3
   W_0^2 W_2}{M^5}\nn\\
   &+\frac{0.050583 \alpha_0  \beta_0  W_0^2
   W_2}{M^5}-\frac{3.07\times10^{-6} \alpha_0  W_0^2
   W_2}{M^5}\nn\\
   &+\frac{0.005757 \gamma_0  W_0^2 W_2}{M^5}-\frac{0.009447 \alpha_0 
   W_4}{M^5}\nn\\
   &-\frac{0.296483 J_1^3}{M^6}-\frac{0.076987 M_2^2}{M^6}+\frac{0.149635 J_1 J_3}{M^6}\nn\\
   &+\frac{0.115867 \alpha_0 ^2 J_1
   W_0 W_2}{M^6}+\frac{0.053359 \beta_0  J_1 W_0
   W_2}{M^6}\nn\\
   &+\frac{0.015886 J_1 W_0 W_2}{M^6}+\frac{0.040706 \alpha_0  M_2
   W_2}{M^6} \nn\\
   &+\frac{0.049323 \alpha_0 
   J_1 M_2 W_0}{M^6}-\frac{0.124683 \alpha_0 ^2
   J_1^2 W_0^2}{M^6}\nn\\
   &-\frac{0.057378 \beta_0  J_1^2
   W_0^2}{M^6}-\frac{0.097335 J_1^2 W_0^2}{M^6}\nn\\
&   -\frac{0.007182 \alpha_0 ^2 W_2^2}{M^6}-\frac{0.000889 \beta_0 
   W_2^2}{M^6}-\frac{1.94\times10^{-11}
   W_2^2}{M^6}\nn\\
   &+\frac{0.037166 \alpha_0  J_1^3
   W_0}{M^7}-\frac{0.568603 J_1^2
   M_2}{M^7}+\frac{0.101511 \alpha_0  J_1^2 W_2}{M^7}\nn\\
   &  -\frac{0.384317 J_1^4}{M^8}  +\ldots \Big)
%
 \end{align}

 \begin{align}  R_{\textrm{{\tiny ISCO}}}=&
   6 M \Big(1+\frac{0.452733 \alpha_0  W_0}{M}  -\frac{0.544331
   J_1}{M^2}\nn\\
&   +\frac{0.040133 \alpha_0 ^2
   W_0^2}{M^2}-\frac{0.053788 \beta_0  W_0^2}{M^2}-\frac{0.059668
   W_0^2}{M^2}\nn\\
   &-\frac{0.110743 \alpha_0  J_1 W_0}{M^3}
   -\frac{0.179922
   M_2}{M^3}-\frac{0.011388 \alpha_0 ^3 W_0^3}{M^3}\nn\\
   &-\frac{0.159112 \alpha_0  \beta_0 
   W_0^3}{M^3}+\frac{0.017135 \alpha_0  W_0^3}{M^3}\nn\\
   &-\frac{0.018819 \gamma_0 
   W_0^3}{M^3}+\frac{0.054768 \alpha_0  W_2}{M^3}\nn\\
   &+\frac{0.054505
   J_3}{M^4}
 -\frac{0.138986 \alpha_0 ^2 J_1
   W_0^2}{M^4}-\frac{0.011282 \beta_0  J_1 W_0^2}{M^4}\nn\\
   &-\frac{0.05465 J_1
   W_0^2}{M^4}
   -\frac{0.226244 J_1^2}{M^4}\nn\\
&    -\frac{0.002465 \alpha_0  M_2
   W_0}{M^4}+\frac{0.001609 \alpha_0 ^4 W_0^4}{M^4}\nn\\
   &-\frac{0.14352 \alpha_0 ^2 \beta_0 
   W_0^4}{M^4}+\frac{0.000237 \alpha_0 ^2 W_0^4}{M^4}\nn\\
   &-\frac{0.023304 \alpha_0  \gamma_0
    W_0^4}{M^4}-\frac{0.015496 \beta_0 ^2 W_0^4}{M^4}\nn\\
    &-\frac{0.000155 \beta_0 
   W_0^4}{M^4}-\frac{0.002151 \delta_0  W_0^4}{M^4}\nn\\
   &-\frac{0.003643
   W_0^4}{M^4}+\frac{0.075878 \alpha_0 ^2 W_0 W_2}{M^4}\nn\\
   &+\frac{0.025303 \beta_0 
   W_0 W_2}{M^4}+\frac{0.00459 W_0 W_2}{M^4}\nn\\
   &-\frac{0.264597 J_1
   M_2}{M^5}-\frac{0.135488 \alpha_0 ^3 J_1 W_0^3}{M^5}\nn\\
   &-\frac{0.105171 \alpha_0  \beta_0  J_1
   W_0^3}{M^5}+\frac{0.022624 \alpha_0  J_1 W_0^3}{M^5}\nn\\
   &-\frac{0.017785 \gamma_0  J_1
   W_0^3}{M^5}+\frac{0.048952 \alpha_0  J_1 W_2}{M^5}\nn\\
&   -\frac{0.075539 \alpha_0 ^2 M_2 W_0^2}{M^5}-\frac{0.025925
   \beta_0  M_2 W_0^2}{M^5}\nn\\
   &-\frac{0.027709 M_2
   W_0^2}{M^5}+\frac{0.016066 M_4}{M^5}\nn\\
   &+\frac{0.059012 \alpha_0 ^3 W_0^2
   W_2}{M^5}+\frac{0.037304 \alpha_0  \beta_0  W_0^2 W_2}{M^5}\nn\\
   &-\frac{0.0006479
   \alpha_0  W_0^2 W_2}{M^5}+\frac{0.004521 \gamma_0  W_0^2
   W_2}{M^5}\nn\\
   &-\frac{0.007425 \alpha_0  W_4}{M^5}+\frac{0.016048 \alpha_0  J_1^2
   W_0}{M^5}\nn\\
   &-\frac{0.021447 \alpha_0  J_3 W_0}{M^5}
   -\frac{0.062736 M_2^2}{M^6}+\frac{0.118328 J_1
   J_3}{M^6}\nn\\
   &+\frac{0.060279 \alpha_0  J_1 M_2 W_0}{M^6}+\frac{0.088564 \alpha_0
   ^2 J_1 W_0 W_2}{M^6}\nn\\
   &+\frac{0.040188 \beta_0  J_1 W_0
   W_2}{M^6}+\frac{0.012688 J_1 W_0 W_2}{M^6}\nn\\
   &+\frac{0.030757 \alpha_0  M_2
   W_2}{M^6}-\frac{0.004997 \alpha_0 ^2 W_2^2}{M^6}\nn\\
   &-\frac{0.000702 \beta_0 
   W_2^2}{M^6}-\frac{1.552\times10^{-11}
   W_2^2}{M^6}\nn\\
   &-\frac{0.230289
   J_1^3}{M^6}-\frac{0.090747 \alpha_0 ^2 J_1^2 W_0^2}{M^6}\nn\\
   &-\frac{0.023107 \beta_0  J_1^2
   W_0^2}{M^6}-\frac{0.077521 J_1^2 W_0^2}{M^6}\nn\\
&   -\frac{0.451308 J_1^2 M_2}{M^7}+\frac{0.068779 \alpha_0  J_1^2
   W_2}{M^7}\nn\\
&   +\frac{0.062687 \alpha_0  J_1^3 W_0}{M^7}
 -\frac{0.297466 J_1^4}{M^8}   
 +\ldots \Big)
 \end{align} 

 \begin{align} 
\Omega_{\textrm{{\tiny ISCO}}}=&\frac{1}{6 \sqrt{6} M}\Big( 1 -\frac{0.75 \alpha_0 
   W_0}{M}+\frac{0.748455
   J_1}{M^2}\nn\\
&   -\frac{0.000218 \alpha_0 ^2 W_0^2}{M^2}+\frac{0.035481 \beta_0 
   W_0^2}{M^2}\nn\\
   &+\frac{0.077688 W_0^2}{M^2}
   +\frac{0.23433 M_2}{M^3}-\frac{0.00784 \alpha_0 ^3
   W_0^3}{M^3}\nn\\
   &+\frac{0.117236 \alpha_0  \beta_0  W_0^3}{M^3}-\frac{0.089082 \alpha_0 
   W_0^3}{M^3}\nn\\
   &+\frac{0.022604 \gamma_0  W_0^3}{M^3}-\frac{0.065506 \alpha_0 
   W_2}{M^3}\nn\\
   &-\frac{0.509757 \alpha_0  J_1 W_0}{M^3}
   +\frac{0.780664 J_1^2}{M^4}-\frac{0.076216   J_3}{M^4}\nn\\
&   -\frac{0.204041 \alpha_0  M_2 W_0}{M^4}-\frac{0.04252 \alpha_0 ^4
   W_0^4}{M^4}\nn\\
   &+\frac{0.009988 \alpha_0 ^2 \beta_0  W_0^4}{M^4}+\frac{0.022952 \alpha_0
   ^2 W_0^4}{M^4}\nn\\
   &+\frac{0.009234 \alpha_0  \gamma_0  W_0^4}{M^4}+\frac{0.017918 \beta_0
   ^2 W_0^4}{M^4}\nn\\
   &+\frac{0.006122\beta_0  W_0^4}{M^4}+\frac{0.002812 \delta_0 
   W_0^4}{M^4}\nn\\
   &+\frac{0.009612 W_0^4}{M^4}-\frac{0.037394 \alpha_0 ^2 W_0
   W_2}{M^4}\nn\\
   &-\frac{0.033029 \beta_0  W_0 W_2}{M^4}-\frac{0.006338 W_0
   W_2}{M^4}\nn\\
   &+\frac{0.114417 \alpha_0 ^2 J_1 W_0^2}{M^4}+\frac{0.075325 \beta_0  J_1 W_0^2}{M^4}\nn\\
   &+\frac{0.168052 J_1
   W_0^2}{M^4}
   +\frac{0.107837 \alpha_0 ^2 M_2 W_0^2}{M^5}\nn\\
   &+\frac{0.051719 \beta_0
    M_2 W_0^2}{M^5}+\frac{0.065906 M_2 W_0^2}{M^5}\nn\\
    &-\frac{0.022183
   M_4}{M^5}+\frac{0.003951 \alpha_0 ^3 W_0^2 W_2}{M^5}\nn\\
   &-\frac{0.024892
   \alpha_0  \beta_0  W_0^2 W_2}{M^5}-\frac{0.001403 \alpha_0  W_0^2
   W_2}{M^5}\nn\\
   &-\frac{0.006125 \gamma_0  W_0^2 W_2}{M^5}+\frac{0.010091 \alpha_0 
   W_4}{M^5}\nn\\
   &-\frac{0.55815 \alpha_0  J_1^2
   W_0}{M^5}+\frac{0.093943 \alpha_0  J_3 W_0}{M^5}\nn\\
   &+\frac{0.646435 J_1
   M_2}{M^5}+\frac{0.04601 \alpha_0 ^3 J_1 W_0^3}{M^5}\nn\\
   &+\frac{0.260446 \alpha_0  \beta_0 
   J_1 W_0^3}{M^5}-\frac{0.190736 \alpha_0  J_1 W_0^3}{M^5}\nn\\&
   +\frac{0.05127 \gamma_0  J_1
   W_0^3}{M^5}-\frac{0.144884 \alpha_0  J_1 W_2}{M^5}\nn\\
&   +\frac{0.12833 M_2^2}{M^6}+\frac{0.981402
   J_1^3}{M^6}+\frac{0.287271 \alpha_0 ^2 J_1^2 W_0^2}{M^6}\nn\\
   &+\frac{0.147587 \beta_0  J_1^2
   W_0^2}{M^6}+\frac{0.315357 J_1^2 W_0^2}{M^6}\nn\\
   &-\frac{0.065011 \alpha_0  M_2
   W_2}{M^6}+\frac{0.009761 \alpha_0 ^2 W_2^2}{M^6}\nn\\
   &+\frac{0.000971 \beta_0 
   W_2^2}{M^6}+\frac{2.328\times10^{-11}
   W_2^2}{M^6}\nn\\
   &-\frac{0.260058 J_1
   J_3}{M^6}-\frac{0.605842 \alpha_0  J_1 M_2 W_0}{M^6}\nn\\
   &-\frac{0.115681 \alpha_0 ^2
   J_1 W_0 W_2}{M^6}-\frac{0.094429 \beta_0  J_1 W_0
   W_2}{M^6}\nn\\
   &-\frac{0.025455 J_1 W_0 W_2}{M^6}
   +\frac{1.42856 J_1^2 M_2}{M^7}\nn\\
   &   -\frac{0.782788 \alpha_0  J_1^3 W_0}{M^7}-\frac{0.276655 \alpha_0  J_1^2
   W_2}{M^7}\nn\\
&
+ 
\frac{1.38245 J_1^4}{M^8}
+\ldots \Big).
 \end{align} 

\bsp \label{lastpage}

\end{document}